\documentclass[aps,superscriptaddress,showpacs,pre,10pt,nofootinbib,twocolumn]{revtex4-2}

\usepackage{booktabs}
\usepackage{bm} 
\usepackage{dsfont}
\usepackage{graphicx} 
\usepackage{amsmath}
\newcommand\norm[1]{\left\lVert#1\right\rVert}
\usepackage{amsthm}
\usepackage{amssymb} 
\usepackage{epstopdf} 
\usepackage{comment}
\usepackage{amsfonts}
\usepackage{epsfig}
\usepackage{color} 
\usepackage[dvipsnames]{xcolor}
\usepackage{subfigure}
\usepackage[english]{babel}
\usepackage[bb=boondox]{mathalfa}
\usepackage{mathtools}
\usepackage{dsfont}
\usepackage{tikz}
\usetikzlibrary{quantikz2}
\usetikzlibrary{arrows.meta,
                chains,
                positioning,
                shapes.geometric,
                shadows,shapes,positioning,trees,
                quantikz2, angles, quotes,
                calc
                }
\tikzset{block/.style={draw,minimum height=2cm,minimum width=2cm}}
\tikzset{block2/.style={draw,minimum width=2cm}}

\usepackage{hyperref}
\usepackage[normalem]{ulem} % For striking out texts!
\hypersetup{
    colorlinks=true,       % false: boxed links; true: colored links
    linkcolor=cyan,          % color of internal links
    citecolor=magenta,        % color of links to bibliography
    filecolor=magenta,      % color of file links
    urlcolor=cyan,           % color of external links
    runcolor=cyan
}
\newcommand{\vcellh}[1]{%
  \parbox[c][0.5cm][c]{\linewidth}{#1}%
}
\newcommand{\vcell}[1]{%
  \parbox[c][1.5cm][c]{\linewidth}{#1}%
}

\newcommand {\be}{\begin{equation}}
\newcommand {\ee}{\end{equation}}

\newcommand{\ba}{\begin{eqnarray}}
\newcommand{\ea}{\end{eqnarray}}

\newcommand{\ignore}[1]{}

\usepackage[margin=1in,nofoot]{geometry}

\newcommand{\rmd}{{\text d}}
\renewcommand{\Re}{\operatorname{Re}}

\newcommand{\beq}{\begin{equation}}
\newcommand{\eeq}{\end{equation}}
\newcommand{\beqnn}{\begin{equation*}}
\newcommand{\eeqnn}{\end{equation*}}
\newcommand{\bea}{\begin{eqnarray}}
\newcommand{\eea}{\end{eqnarray}}
\newcommand{\beann}{\begin{eqnarray*}}
\newcommand{\eeann}{\end{eqnarray*}}
\newcommand{\bes} {\begin{subequations}}
\newcommand{\ees} {\end{subequations}}

\newcommand{\bu} {{\bf u}}

\begin{document}
\raggedbottom 
\title{Quantum algorithm for differential equations via \\
permutation matrix representation
with application to the Burgers' equation}
\author{Hriday Sabharwal}
\email{hsabharw@usc.edu}
\affiliation{Department of Physics and Astronomy, University of Southern California, Los Angeles, California 90089, USA}
\author{Amir Kalev}
\affiliation{Department of Physics and Astronomy, University of Southern California, Los Angeles, California 90089, USA}
\affiliation{Information Sciences Institute, University of Southern California,
Marina del Rey, California 90292, USA}
\author{Itay Hen}
%\email{itayhen@isi.edu}
\affiliation{Department of Physics and Astronomy, University of Southern California, Los Angeles, California 90089, USA}
\affiliation{Information Sciences Institute, University of Southern California, Marina del Rey, California 90292, USA}
\affiliation{Department of Electrical and Computer Engineering, University of Southern California, Los Angeles, California 90089, USA}

\date{\today}

\begin{abstract}
\noindent 
We develop a quantum algorithm for solving the dynamics of the nonlinear viscous Burgers’ equation. We apply the Carleman linearization procedure on the spatially discretized equation, followed by a padding scheme that allows implementation on qubit registers. Existing Carleman-based quantum algorithms commonly formulate the lifted linear differential equation in an oracle model. Here we decompose the padded generator into diagonal masks and reversible arithmetic permutations using the Permutation Matrix Representation (PMR), which we show to be naturally compatible with the Linear Combination of Hamiltonian Simulations (LCHS) algorithm. Under the assumptions required by LCHS—most importantly positive semidefiniteness of the Hermitian part of the linear generator, possibly after a stabilizing shift—the algorithm prepares a normalized quantum state proportional to the solution of the truncated lifted system; the stabilizing shift introduces an exponential postselection overhead, which we quantify and mitigate through a rescaling scheme. We show that our algorithm scales with the off-diagonal norm of the Carleman generator instead of the matrix norm, which can be advantageous for other generators that are diagonally dominant. We also extend the PMR scheme to general fluid equations that may contain higher-order derivatives or nonlinear terms, or may involve multiple fluid variables or spatial dimensions. The construction illustrates how PMR can serve as a convenient Hamiltonian-simulation primitive for a broader class of LCU-based algorithms.
\end{abstract}

\maketitle

\section{Introduction}

Nonlinear differential equations appear throughout the natural and social sciences---in physics, engineering, chemistry, biology, and economics, among other fields---and are a workhorse of applied mathematics. Solving them numerically on classical computers typically requires either spatial discretization or some form of linearization, both of which substantially enlarge the dimension of the underlying problem. As a result, accurate numerical solution of large nonlinear systems remains computationally demanding.

Quantum computers offer a promising avenue for tackling nonlinear dynamics. Several distinct strategies have been pursued in the literature. Hybrid quantum-classical methods~\cite{Lubasch_2020,Kyriienko_2021,SHUKLA2023127708} reformulate the differential equation as an optimization problem, which is then solved using a parametrized quantum circuit. Such schemes are well suited to near-term hardware, but no provable scaling advantage is currently known. A second class of approaches is based on mean-field evolution of many copies of a quantum system~\cite{lloyd2020quantumalgorithmnonlineardifferential}, which produces effective nonlinear dynamics on subsystems and is expected to scale favorably, though it is difficult to realize on near-term devices. A third strategy uses linearization techniques to embed the nonlinear equation into a higher-dimensional linear system~\cite{Liu_2021,Krovi_2023} via  Carleman linearization~\cite{Carleman_1932}; the resulting system can then be addressed using established quantum simulation algorithms for linear ODEs~\cite{Harrow_2009,Childs_2017}.

In this work we propose an explicit quantum implementation for the spatially discretized and Carleman linearized viscous Burgers' equation. We provide a padding scheme that enables the implementation of the matrix on qubit registers, following which we decompose the Carleman generator using the Permutation Matrix Representation (PMR) in terms of quantum arithmetic primitives such as spatial shifts and adder/subtractors. We then propose a general algorithm that employs the PMR decomposition to solve the linear system by combining the PMR algorithm for Hamiltonian simulation~\cite{Kalev_2021, Kalev_2025} and the Linear Combination of Hamiltonian Simulations (LCHS) algorithm for system of linear ODEs~\cite{An_2023, An_2025}. While previous work~\cite{Liu_2021,Krovi_2023} provides rigorous Carleman convergence analysis that we exclude from the scope of this paper, our method provides an explicit gate construction for the Carleman matrix instead of relying on oracle access. Our method also shows scaling advantages for equations where the Carleman generator is dominated by the diagonal part. We also provide a blueprint that extends the PMR scheme to general nonlinear fluid equations. Our results as compared to other algorithms are summarized in Table~\ref{tab:solver-comparison}.

\begin{table*}[t]
\centering
\renewcommand{\arraystretch}{1.25}
\setlength{\tabcolsep}{5pt}

\begin{tabular}{|
  p{0.15\textwidth}|
  p{0.22\textwidth}|
  p{0.2\textwidth}|
  p{0.35\textwidth}|}
\hline\hline
\vcellh{\textbf{Method}} &
\vcellh{\textbf{Access model}} &
\vcellh{\textbf{Normalization}} &
\vcellh{\textbf{Complexity}} \\
\hline

\vcell{LCHS--QSP ~\cite{An_2023,An_2025}} &
\vcell{Block encoding of $X$} &
\vcell{$\alpha_{\mathrm{X}}$\par} &
\vcell{$\tilde{\mathcal{O}}\left(\frac{\norm{y_0}}{\norm{y(t)}}\alpha_Xt(\log(1/\epsilon))^{1/\beta}\right)$ \newline queries to $X$} \\
\hline

\vcell{LCHS--PMR\newline (This paper, Sec.~\ref{sec:lchspmr})} &
\vcell{Permutation and diagonal-mask circuits} &
\vcell{$\Gamma_X$\par} &
\vcell{$\tilde{\mathcal{O}}\left(\frac{\norm{y_0}}{\norm{y(t)}}t\Gamma_XM'\left(\log\left(\frac{1}{\epsilon}\right)\right)^{1+\frac{1}{\beta}}\right)$ gates} \\
\hline\hline

\vcell{Carleman--QLSA ~\cite{Liu_2021,Krovi_2023}} &
\vcell{Sparse or block access to the Carleman matrix} &
\vcell{Norm of the Carleman matrix} &
\vcell{$\tilde{\mathcal{O}}\left(\frac{\norm{u_0}}{\norm{u(t)}}\alpha_Xt\ \text{poly}(L,s,\log(1/\epsilon))\right)$ queries to the block matrices} \\
\hline

\vcell{Carleman--LCHS--PMR\newline (This paper, Sec.~\ref{sec:burgersresources})} &
\vcell{Cyclic shifts, adders/subtractors, register shifts} &
\vcell{Off-diagonal norm of the Carleman matrix} &
\vcell{$\tilde{\mathcal{O}}\!\left(tL^4N^2\frac{\nu}{\zeta^2}\left(\log\!\left(\frac{1}{\epsilon}\right)\right)^{1+\frac{1}{\beta}}\frac{e^{Lt/a\sqrt2}\norm{\tilde\bu_0}}{\norm{\tilde\bu(t)}}\right)$ gates and arithmetic operations} \\
\hline
\end{tabular}

\caption{\label{tab:solver-comparison} Comparison of implementations for linear ODE solvers (LCHS-QSP and LCHS-PMR) and for the nonlinear solvers (Carleman-QLSA and Carleman-LCHS-PMR) applied to the Burgers' equation in the diffusion-dominated regime.
Here, $s$ is the sparsity of the Carleman matrix. For LCHS--QSP we quote the time-independent result of Ref.~\cite{An_2025}, whose QSP-based implementation achieves the exponent $1/\beta$; the time-dependent (truncated-Dyson) case instead carries the exponent $1+1/\beta$. The soft-$\tilde{\mathcal{O}}$ notation absorbs polylogarithmic factors, LCHS state preparation costs and polynomial bit-precision factors.}

\end{table*}

The remainder of this paper is organized as follows. Section~\ref{sec2} reviews the LCHS and PMR algorithms. Section~\ref{sec3} describes the full algorithm specialized to the Burgers' equation, developing the linearization and padding scheme, working out the PMR expansion of the resulting block matrix, concluding with the general PMR-based implementation of the Hamiltonian simulation subroutine inside LCHS. Section~\ref{sec5} analyzes the resource cost, both in general and as specialized to the Burgers' equation. Section~\ref{sec6} demonstrates how the PMR scheme used for the Burgers' equation can be extended to more complex fluid equations. We conclude in Section~\ref{sec7} with a summary and discussion on the broader impact of the PMR method.

\section{Preliminaries}\label{sec2}

In this section we briefly review two essential components of our construction and fix
the notation used in the remainder of the paper. The first is the LCHS
method of Refs.~\cite{An_2023,An_2025}. The second is the PMR approach to Hamiltonian simulation~\cite{Kalev_2021, Kalev_2025}, which acts on a decomposition of the Hamiltonian into diagonal operators and permutations and supplies the divided-difference expansion on which our implementation rests. 

\subsection{LCHS}\label{sec:lchs}

The LCHS construction of Ref.~\cite{An_2023, An_2025} expresses the propagator as a linear combination of unitary time-evolution operators, which can then be implemented through LCU~\cite{Childs_2012,Berry_2015}. The below review follows Ref.~\cite{An_2025}.

Consider a linear, homogeneous, time independent system of ordinary differential equations
\begin{equation}\label{eq:lchsODE}
    \dot{y}=-Xy, \ \ \ y(0)=y_0,
\end{equation}
with $X\in\mathbb{C}^{N\times N}$ and $y\in\mathbb{C}^N$. The solution of Eq.~(\ref{eq:lchsODE}) admits the integral representation
\begin{equation}
    y(t)=e^{-Xt}y_0,
\end{equation}
where the operator $e^{-X(t-t_0)}$ is the propagator. 

The starting point is the integral identity (Theorem~6 of Ref.~\cite{An_2025}), requiring a complex kernel function $f(k)$ and the decomposition of $X$ to the Hermitian matrices $G=\frac{X+X^\dagger}{2}$ and $H=\frac{X-X^\dagger}{2i}$ so that $X=G+iH$. The identity requires that the Hermitian part $G$ be positive semidefinite, $G\succeq 0$~\cite{An_2025}; we will return to this assumption when specializing to the Carleman-truncated Burgers generator.

To implement this on a quantum computer the integral is truncated to $[-K,K]$, and discretized by Gaussian quadrature, giving
\begin{equation}\label{eq:4}
\begin{aligned}
    e^{-Xt}\approx\int_{-K}^Kg(&k)U(t,k)\mathrm{d}k\\&=\sum_{m=-K/h_1}^{K/h_1-1}\int_{mh_1}^{(m+1)h_1}g(k)U(t,k)\mathrm{d}k
    \\&\approx\sum_{m=-K/h_1}^{K/h_1-1}\sum_{q=0}^{Q_{\rm GQ}-1}c_{q,m}U(t,k_{q,m}),
\end{aligned}
\end{equation}
where $g(k)=\frac{f(k)}{1-ik}$, $U(t,k)=e^{-it(kG+H)}$, $h_1$ is the step size, $Q_{\rm GQ}$ is the number of Gaussian nodes per subinterval, $k_{q,m}$ are the nodes themselves, and $c_{q,m}=\frac{h_1}{2}w_q\,g(k_{q,m})$ with $w_q$ the corresponding Gaussian weights. The result is an LCU with $J:=2KQ_{\rm GQ}/h_1$ terms. The parameters $K$, $h_1$, and $Q_{\rm GQ}$---which we will eventually condense into $K$ and $J$---control the accuracy of the approximation and ultimately determine the resource cost of the algorithm. Relabeling, we write
\begin{equation}\label{eq:lcusum}
    \sum_{j=0}^{J-1}c_jU(t,k_j).
\end{equation}
where $\norm{c}_1=\sum_j|c_j|=\Theta(1)$. We emphasize that the nodes $k_j$ are the union of Gaussian nodes across all subintervals; in particular, they are not uniformly spaced on $[-K,K]$.

We fix the optimal kernel function demonstrated in Ref.~\cite{An_2025}, controlled by the parameter $\beta\in(0,1)$. We also note that the inhomogeneous and time dependent cases can be treated in a similar manner.

We note that because LCU is being used to implement a non-unitary operator, oblivious amplitude amplification is not directly applicable to the outer LCHS combination; its success probability is instead boosted by standard amplitude amplification. Oblivious amplitude amplification is, however, still employed within each (unitary) Hamiltonian-simulation segment, cf.~Sec.~\ref{gatecost}.

\subsection{PMR}\label{sec:pmr}

The Permutation Matrix Representation (PMR) approach to Hamiltonian simulation~\cite{Kalev_2021,Kalev_2025} separates the diagonal and off-diagonal pieces of the generator and resums the diagonal evolution exactly through divided differences. The exposition below follows Ref.~\cite{Kalev_2025}.

The PMR form of a Hamiltonian is
\begin{equation}\label{eq:pmrform}
H = D_0 + \sum_{i=1}^M D_i P_i,
\end{equation}
where $D_i$ are (generally complex valued) diagonal operators and $P_i$ are off-diagonal permutation operators i.e. they do not have any diagonal elements. This decomposition is the foundation for an expansion of the evolution operator in terms of divided differences~\cite{Kalev_2025}. To optimize resources, the total simulation time is divided into $r$ intervals of duration $\Delta t = t/r$~\cite{Kalev_2025}. Upon inserting the decomposition into a Taylor expansion of the time evolution operator, and after some algebra~\cite{Albash_2017,Gupta_2020}, the expansion reorganizes into walks on the basis states, with the diagonal evolution along each walk resummed by a divided difference of the exponential:
\begin{equation}\label{eq:pmrwalk}
U(\Delta t) = \sum_{q=0}^\infty \frac{\Delta t^q}{q!}\sum_{\mathbf{i}_q}\Gamma_{\mathbf{i}_q} P_{\mathbf{i}_q} A_{\mathbf{i}_q},
\end{equation}
where $\mathbf{i}_q = (i_1, i_2, \ldots, i_q)$ is a tuple of $q$ indices each ranging from $1$ to $M$, and $P_{\mathbf{i}_q} := P_{i_q}\cdots P_{i_2} P_{i_1}$ is the ordered product of off-diagonal permutations along the walk. The diagonal operators $A_{\mathbf{i}_q}$ are
\begin{equation}\label{eq:Aiq}
A_{\mathbf{i}_q} = \sum_z \frac{d_{\mathbf{i}_q}}{\Gamma_{\mathbf{i}_q}}\frac{q!}{\Delta t^q} e^{-i\Delta t[E_{z_0},E_{z_1},\ldots,E_{z_q}]} |z\rangle\langle z|,
\end{equation}
where $e^{-i\Delta t[E_{z_0},E_{z_1},\ldots,E_{z_q}]}$ is the divided difference of the exponential evaluated at the diagonal energies $E_{z_j} = \langle z_j | D_0 | z_j \rangle$ along the walk $|z_j\rangle = P_{i_j}\cdots P_{i_1}|z\rangle$ (with the convention $z_0 := z$). The walk weights are $d_{\mathbf{i}_q} = \prod_{j=1}^q d_{i_j}(z_j)$ with $d_{i_j} = \langle z_j | D_{i_j}| z_j \rangle$, and $\Gamma_{\mathbf{i}_q} = \prod_j \Gamma_{i_j}$ with $\Gamma_i \geq \max_z |d_i(z)|$.

The first approximation in this algorithm replaces each divided-difference exponential by a small linear combination of phases. For sufficiently small $\delta$ one has~\cite{Kalev_2025}
\begin{equation}\label{eq:ddapprox}
e^{-i\delta[x_0,\ldots,x_q]} \approx \frac{(-i\delta)^q}{q!} e^{-i\delta \frac{1}{q+1}\sum_{j=0}^q x_j}.
\end{equation}
Combining this approximation with $K'$ successive applications of the Leibniz rule for divided differences~\cite{whittaker1944calculus,deboor2005divideddifferences}, followed by the algebraic manipulations of Ref.~\cite{Kalev_2025}, yields the representation
\begin{equation}\label{eq:Uhat}
\hat{U} = \sum_{q=0}^\infty \sum_{\mathbf{i}_q}\sum_{\mathbf{k}_q} \frac{\Gamma_{\mathbf{i}_q}\Delta t^q}{(K')^q q!}V_{(\mathbf{i}_q,\mathbf{k}_q)},
\end{equation}
with
\begin{equation}\label{eq:V}
V_{(\mathbf{i}_q,\mathbf{k}_q)} = (-i)^q P_{\mathbf{i}_q} \sum_z e^{-i\delta \sum_{s=0}^q \alpha_s E_{z_s}}|z\rangle\langle z|,
\end{equation}
and where $\mathbf{k}_q = (k_1, \ldots, k_q)$ is a tuple of $q$ indices each ranging from $1$ to $K'$, and $\delta := \Delta t/K'$. The relation between $j_1, \ldots, j_{K'}$ and $k_1, \ldots, k_q$ is that $j_l$ is the number of indices in $\mathbf{k}_q$ whose value is $l$. The coefficients $\alpha_s$ are defined by the procedure in Ref.~\cite{Kalev_2025} and depend on $\mathbf{k}_q$.

The error of this first approximation is controlled by $K'$, the number of subdivisions of the divided-difference exponential generated by the Leibniz rule.

Equation~(\ref{eq:Uhat}) is now a linear combination of unitaries, but the sum over $q$ is infinite and must be truncated for an LCU implementation. The second approximation is the truncation order $Q$:
\begin{equation}\label{eq:Utilde}
\hat{U} \approx \tilde{U} := \sum_{q=0}^Q \frac{(\Gamma\Delta t)^q}{q!}\sum_{\mathbf{i}_q}\frac{\Gamma_{\mathbf{i}_q}}{\Gamma^q}\sum_{\mathbf{k}_q}\frac{1}{(K')^q}V_{(\mathbf{i}_q,\mathbf{k}_q)},
\end{equation}
where $\Gamma=\sum_i\Gamma_i$ is the off-diagonal norm. This approximation is implemented through the standard LCU procedure~\cite{Berry_2015,Childs_2012}, and $\tilde{U}$ is then applied $r$ times to produce the final approximation to the long-time evolution.

\section{Quantum Algorithm for Burgers' equation}\label{sec3}

In this section we provide an explicit quantum algorithm that prepares a quantum state encoding an approximate solution to the Burgers' equation. 

The equation is first discretized and linearized through the Carleman linearization procedure, and the resulting matrix is decomposed using PMR in terms of simple arithmetic operations. This enables the use of the PMR Hamiltonian simulation algorithm of Sec.~\ref{sec:pmr} in the Hamiltonian simulation subroutine of the LCHS algorithm of Sec.~\ref{sec:lchs} as a combined quantum implementation.

\begin{figure*}[t]
\centering
    \includegraphics[width=1.0\linewidth]{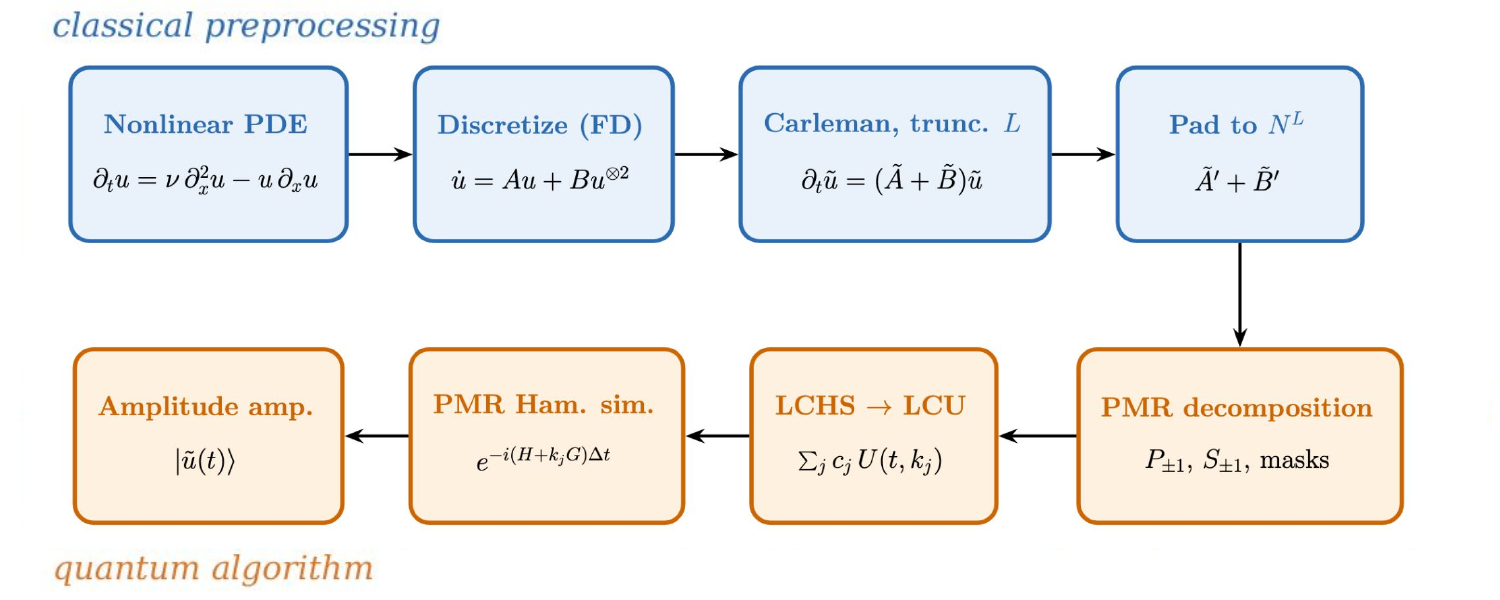}
	\caption{Overview of the LCHS--PMR construction for a nonlinear fluid equation. In the classical preprocessing stage, the PDE is discretized in space, lifted to a finite linear system by Carleman linearization truncated at level $L$, and padded to a power-of-two block dimension $N^L$. The quantum algorithm then decomposes the padded generator $\tilde{A}'+\tilde{B}'$ into PMR primitives---shift operators $P_{\pm1}$, modular adders and subtractors $S_{\pm1}$, and diagonal masks---expresses the non-unitary propagator through the LCHS integral as a linear combination of unitaries over quadrature nodes $k_j$, simulates each $H+k_jG$ by PMR, and recovers the normalized solution state $\ket{\tilde{\bu}(t)}$ after amplitude amplification.}
\label{figpipeline}
\end{figure*}

The overall construction, from the nonlinear PDE through Carleman linearization and PMR decomposition to the LCHS implementation, is summarized in Fig.~\ref{figpipeline}.

\subsection{Discretization, Carleman linearization and padding of the Burgers' equation}\label{sec:carleman}

The viscous Burgers' equation is a nonlinear partial differential equation that captures the interplay of nonlinear advection and diffusive smoothing in a one-dimensional flow:

\begin{equation}
    \frac{\partial u}{\partial t}=\nu\frac{\partial^2u}{\partial x^2}-u\frac{\partial u}{\partial x}.
\end{equation}
Here $u(x,t)$ is a velocity-like field; the term $u\,\partial u/\partial x$ encodes self-advection (waves moving at speeds proportional to their own amplitude), and $\nu\,\partial^2 u/\partial x^2$ is the dissipative term that smooths gradients. The equation may be viewed as a tractable analogue of the Navier--Stokes equations: it retains the essential nonlinearity and dissipative structure but admits a closed-form solution via the Cole--Hopf transformation~\cite{Cole_1951,Hopf_1950}. The competition between steepening and diffusion makes Burgers' equation a standard testbed for shock formation, turbulence-like dynamics, and transport phenomena in fluid mechanics, acoustics, and traffic flow~\cite{Bagheri_2018}.

Throughout the remainder of the paper we work in nondimensional variables: lengths are measured in units of the domain size, and velocities in units of a characteristic velocity $U_\star$ of the initial data, which we take to be $U_\star=\max_x|u_0(x)|$ (times are then measured in units of length over $U_\star$). In these units $u$, $a$, $\zeta$, and $t$ are dimensionless, $\|u_0\|_\infty\leq 1$, and $\nu$ denotes the dimensionless viscosity, i.e.\ the inverse Reynolds number associated with the scales above. In particular, quantities such as $\nu/a^2$ and $1/a$ appearing below are directly comparable, and regime conditions like $\nu N/\zeta\gtrless 1$ are statements about the cell Reynolds number $U_\star a/\nu$ of the physical flow. We note that all gate costs are computed in terms of the dimensionless quantities, which in turn depend on the characteristic velocity and domain size. We also note the additional discussion in Appendix~\ref{a4} on the rescaling of $u(t)$, that also follows the dimensionless scheme.

The equation must first be brought to the linear form of Eq.~\eqref{eq:lchsODE}. We do this through Carleman linearization~\cite{Carleman_1932,Liu_2021}, with a slight modification (introduced below) that adapts the resulting block structure to qubit registers. For clarity we restrict the spatial domain to one dimension; the construction generalizes straightforwardly to higher dimensions.

We discretize the spatial dimension first, replacing $u(x,t)$ by functions $u_j(t)$, $j=0,1,\dots,N-1$, corresponding to grid points $x_j=ja$ with spacing $a$ on a periodic domain of total length $\zeta=Na$. Periodic boundary conditions, $u_{j+N}\equiv u_j$, are imposed throughout. When the physical problem of interest is posed on the whole line with localized (decaying) initial data, the periodic box is taken large enough that $u_j\approx 0$ near the boundaries for the duration of the simulation, so that periodicity introduces no spurious wraparound effects; for genuinely periodic data no such condition is needed. The spatial derivatives become finite differences,
\beq
\frac{\partial u}{\partial x}   \to  \frac{1}{2 a} \left( u_{j+1} - u_{j-1} \right) \,,
\eeq
\beq
\frac{\partial^2 u}{\partial x^2}   \to  \frac{1}{a^2} \left( u_{j+1} -2 u_j +u_{j-1} \right)  \,,
\eeq
where the index arithmetic is understood mod $N$. Higher-order finite-difference stencils may be used, with the usual improvement in spatial truncation error at the cost of additional shift terms in the PMR decomposition. Substituting these into Burgers' equation gives, for each $j$,
\beq
\dot{u}_j = \frac{\nu}{a^2} \left( u_{j+1} -2 u_j +u_{j-1} \right)-  \frac{1}{2 a} u_j \left( u_{j+1} - u_{j-1} \right) \,. 
\eeq
Stacking the $u_j(t)$ into a column vector $\bu = (u_0,\ldots,u_{N-1})$, the discretized equation takes the compact form
\beq\label{eq:discreteburger}
\dot{\bu} =  A \bu  + B \bu^{\otimes 2}  \,,
\eeq
with the matrices $A$ and $B$ specified below.

Let us denote by $\mathbf{e}_j:=(0,\ldots,0,1,0,\ldots,0)$ the $j$-th standard basis vector in the space of $\bu$. The matrix $A$ has zeros except for $A_{jj} = -2 \nu / a^2$ and $A_{j,j\pm1} =  \nu / a^2$ (with index arithmetic mod $N$), and admits the PMR form $A = (-2 \nu / a^2) P_0 + ( \nu / a^2) P_1+( \nu / a^2) P_{-1}$ in terms of the cyclic shift operators
\begin{equation}\label{eq:Pshift}
    P_j:=\sum_{k=0}^{N-1}\mathbf{e}_k\cdot \mathbf{e}_{k\oplus j}^T, \ \ \ \ \ \ P_0=\mathds{1},
\end{equation}
where $\oplus$ denotes addition mod $N$. The nonlinear matrix $B$ acts on $\bu^{\otimes 2}$ and produces an $N$-vector; its only nonzero entries satisfy $B_{[j,(j,j\oplus 1)]}=-B_{[j,(j,j\ominus 1)]}=-1/(2a)$, so that $B=-1/(2a) Q_1 + 1/(2a)Q_{-1}$ with
\begin{equation}\label{eq:Qpm1}
    Q_j:=\sum_{k=0}^{N-1}\mathbf{e}_k\cdot(\mathbf{e}_k\otimes\mathbf{e}_{k\oplus j})^T.
\end{equation}
The matrix $B$ thus maps vectors of the dimension of $\bu^{\otimes 2}$ to vectors of the dimension of $\bu$.

To linearize the system we follow Carleman's strategy (derived in detail in Appendix~\ref{a:carleman}) and generalize Eq.~\eqref{eq:discreteburger} to higher-order tensor powers $\bu^{\otimes k}$: 

\be
\frac{\rmd \bu^{\otimes k}}{\rmd t}=A_k \bu^{\otimes k}  + B_k \bu^{\otimes (k+1)},
\ee

where
\beq\label{eq:AkBk}
A_k = \sum_{j=1}^k A^{(j,k)}
\quad
\textrm{and} 
\quad
 B_k = \sum_{j=1}^k B^{(j,k)},
\eeq
and where
\begin{equation}
    A^{(j,k)}:=\mathds{1} \otimes \cdots \otimes A \otimes \ldots \otimes  \mathds{1}
\end{equation}
denotes $A$ tensored with $k-1$ identities, with $A$ in the $j$-th position. The same convention is used for $B^{(j,k)}$.

These equations form an infinite hierarchy: the dynamics of $\bu^{\otimes k}$ couples to $\bu^{\otimes (k+1)}$ through $B_k$. To obtain a finite linear system we truncate the hierarchy at some level $L$, retaining $\bu^{\otimes 1},\ldots,\bu^{\otimes L}$ and discarding the coupling $B_L\bu^{\otimes(L+1)}$. 

\begin{figure*}[htb]
\centering
	\includegraphics[width=0.85\linewidth]{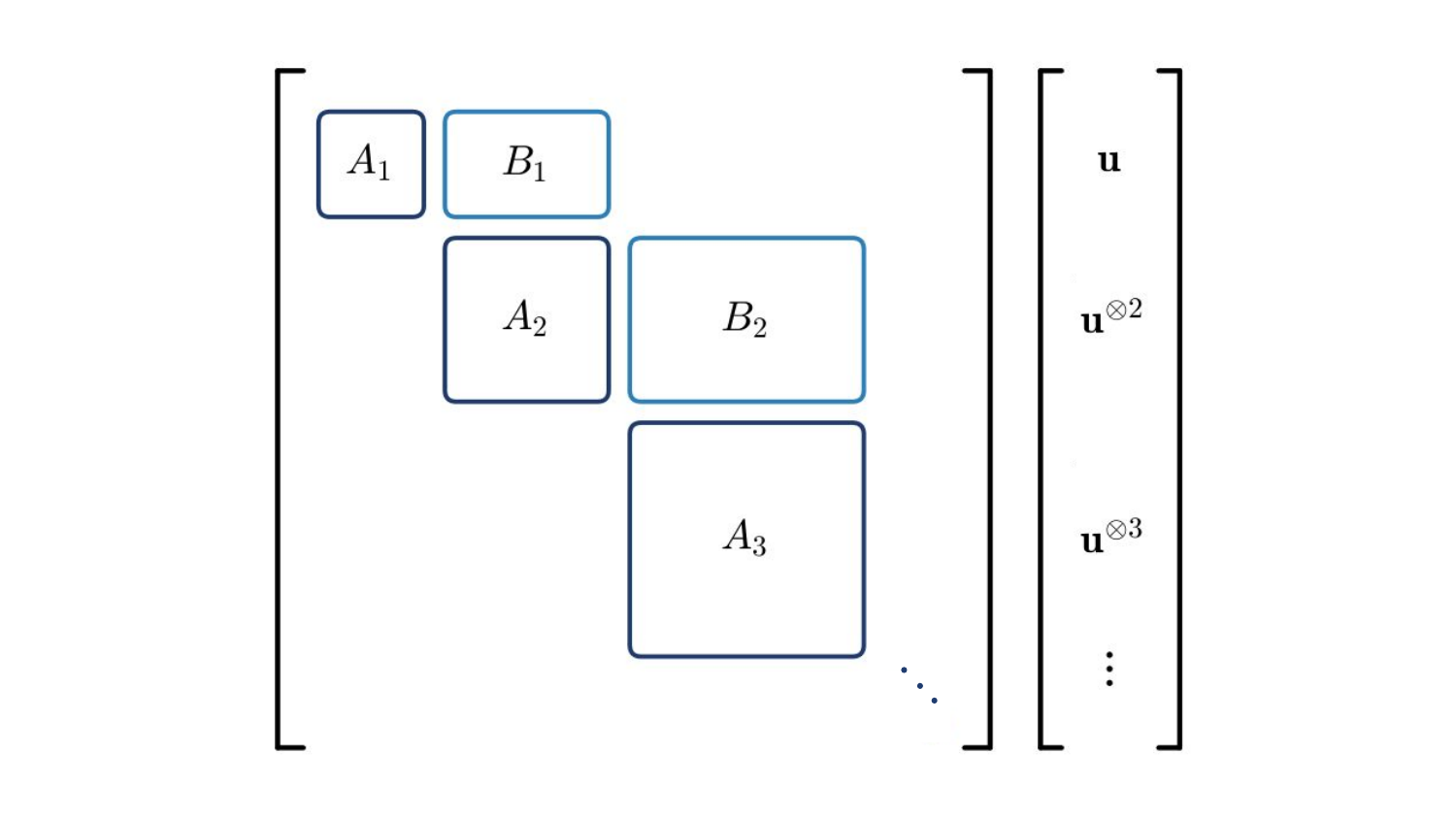}
	\caption{Structure of the matrix $(\tilde{A} + \tilde{B})$ in the extended space (left) acting on the solution vector (right). The diagonal blocks $A_k$ implement the linear part of the dynamics within the $\bu^{\otimes k}$ sector, while the off-diagonal blocks $B_k$ couple $\bu^{\otimes k}$ to $\bu^{\otimes(k+1)}$.}
\label{figblock}
\end{figure*}

It is convenient to view the $A_k$ and $B_k$ as block matrices in the extended Hilbert space spanned by the basis vectors of $\bu \cup \bu^{\otimes 2} \cup \cdots \cup \bu^{\otimes L}$. Define the block vector
\beq
\tilde{\bu}:=\bigoplus_{k=1}^L\bu^{\otimes k},
\eeq
which assembles the truncated hierarchy into one big vector. In this representation the coupled system takes the closed linear form
\beq\label{eq47}
\dot{\tilde{\bu}} = (\tilde{A} + \tilde{B})  \tilde{\bu},
\eeq
where $\tilde{A}$ contains the diagonal blocks $A_1,...,A_L$ and $\tilde{B}$ contains the off-diagonal blocks $B_1,...,B_{L-1}$ that couple consecutive levels of the hierarchy, as illustrated in Fig.~\ref{figblock}. 

To implement Eq.~(\ref{eq47}) on a quantum computer, the matrices and vectors must have dimensions that are powers of two. The block matrix $\tilde A+\tilde B$ and the block vector $\tilde\bu$ have size $\sum_{k=1}^LN^k$, which is not generally a power of two. We resolve this with a small modification to the construction.

Choose $N$ to be a power of two, $N=2^n$, and pad each block vector $\bu^{\otimes k}$ with zeros so that all blocks have the same size $N^L$. Physically, this corresponds to introducing $L-k$ extra qubits in the $\ket 0$ state for each $\ket{\bu}^{\otimes k}$, producing $\ket{\bu}^{\otimes k}\ket{0}^{\otimes L-k}$ (with each ket written in the $N$-dimensional computational basis). Once all blocks share the same size, the full state can be written as a tensor product of an outer label register with the padded contents:
\begin{equation}\label{eq:utildestate}
    \ket{\tilde\bu}\;=\;\frac{1}{\norm{\tilde\bu}}\sum_{k=1}^{L}\norm{\bu}^k\,\ket{k-1}\otimes \left(\ket{\bu}^{\otimes k}\otimes\ket0^{\otimes L-k}\right),
\end{equation}
where $\ket{k}$ lives in the $L$-dimensional outer register (which will itself be chosen of dimension a small power of two), $\ket\bu:=\bu/\norm\bu$ is the normalized amplitude encoding of $\bu$, namely $\norm{\tilde\bu}^2=\sum_{k=1}^L\norm{\bu}^{2k}$, and the prefactors $\norm{\bu}^k$ ensure that the encoded state agrees with the lifted Carleman vector $\tilde\bu=\bigoplus_{k=1}^L\bu^{\otimes k}$ up to overall normalization. The cost of preparing the input state $\ket{\tilde\bu_0}$ at $t=0$ is treated as an oracle assumption: we assume access to a unitary that prepares $\ket{\tilde\bu_0}$ from a reference state, and absorb its complexity into the soft-$\tilde{\mathcal{O}}$ in resource estimates.

Likewise, the matrix $\tilde{A}+\tilde{B}$ also needs to be padded so that each of its blocks has the common size $N^L\times N^L$. We choose the following paddings
\be\label{eq:Apadding}
A_k':=A_k\otimes\mathds1^{\otimes L-k},
\ee
\be\label{eq:Bpadding}
B_k':=B_k\otimes(\ket0\otimes\mathds1^{\otimes L-k-1}),
\ee
which retain the structure of Eq.~\eqref{eq47} in the non-padding registers. The benefit of equal block sizes is that the full padded matrix can be written as a tensor product over the outer $L$-dimensional register, which is structurally simple. The following subsections describe the padding scheme so that the resulting PMR expansion uses only standard arithmetic permutations and diagonal masks.

\subsection{PMR decomposition of the generator} 

Here we define the arithmetic permutations required to build the PMR expansion for the padded Carleman matrix. We state the final expansions here and give a full derivation in Appendix~\ref{a:pmr}.

Recall the PMR expansion of $A$ from the previous section,

\begin{equation}\label{eq:Apmr}
    A=(-2\nu/a^2)\mathds{1}+(\nu/a^2)P_1+(\nu/a^2)P_{-1}.
\end{equation}
There are two indices which will control the position of $A$ in the full tensor product; $k$ labels the Carleman subspace and $j$ labels the position of $A$ within the tensor product of that subspace i.e. $A^{(j,k)}$. If we expand $A$ using Eq.~\eqref{eq:Apmr} within these labels, each $j$ and $k$ will define a different permutation operator, thus the size of the expansion should run through all labels $j=(1,...,k)$ and $k=(1,...,L)$.

However, the padding of Eq.~\eqref{eq:Apadding} helps unify the permutations and prompts us to collect the diagonals, significantly reducing the size. On top of that, it allows us to assign blocks to each $A_k'$ by tensoring with the $L$ dimensional matrix $\Lambda_k:=\text{diag}(0,...,1,...,0)$ with the single 1 at position $(k,k)$.

The final PMR expansion for the full padded block-diagonal matrix $\tilde A'$ takes the form
\begin{equation}
\begin{aligned}
    \tilde{A}'&=\frac{-2\nu}{a^2}\sum_{k=1}^Lk\left(\Lambda_k\otimes\mathds{1}^{\otimes L}\right)\\&\quad+\frac{\nu}{a^2}\sum_{x=\pm1}\sum_{j=1}^L(\Lambda^{(j)}\otimes\mathds{1}^{\otimes L})(\mathds{1}_L\otimes P_x^{(j,L)}),
\end{aligned}
\end{equation}
where $\Lambda^{(j)}=\sum_{k=j}^L\Lambda_k$,
$\mathds{1}_L$ is the $L\times L$ identity, $\mathds{1}^{\otimes L}$ is the $N^L\times N^L$ identity, and $P_{\pm1}^{(j,L)}$ denotes $P_{\pm1}$ tensored with $L-1$ identities, with $P_{\pm1}$ at the $j$-th position. Each $P_{\pm1}^{(j,L)}$ is itself a permutation matrix. The derivation can be found in Appendix~\ref{a:pmr}.

The construction for $\tilde B$ is similar in spirit but slightly different in two aspects. Firstly, each $B_k$ is rectangular: it maps $N^{k+1}$-dimensional vectors to $N^k$-dimensional vectors. The padding and PMR for $B_k$ is thus not as trivial as $A_k$. Second, each $B_k$ belongs to off-diagonal blocks, requiring the use of shift permutations on the $L$ dimensional register. 

The padding of Eq.~\eqref{eq:Bpadding} can be better understood as two combined paddings. The first, $B_k\otimes \ket{0}$, pads the rectangular matrix to a square matrix, which can be written as a PMR. The second, $(B_k\otimes\ket0)\otimes\mathds1^{\otimes L-k-1}$, pads the $k$-th Carleman subspace to a common overall dimension. We can thus derive a PMR completely with the first padding, and the second padding carries over naturally.

Recall the PMR expansion of $B$:

\begin{equation}\label{eq:Bpmr}
    B=(-1/2a)Q_1+(1/2a)Q_{-1}.
\end{equation}
The same two indices $j$ and $k$ control the size of the PMR here as well. In the $N$-dimensional computational basis, the rectangular $N\times N^2$ matrices $Q_1$ and $Q_{-1}$, with periodic boundary conditions, read
\begin{equation}\label{eq:Qexplicit}
    Q_1=\sum_{i=0}^{N-1}\ket{i}\bra{i,i\oplus 1},\quad Q_{-1}=\sum_{i=0}^{N-1}\ket{i}\bra{i,i\ominus 1},
\end{equation}
where $\oplus$ and $\ominus$ denote addition and subtraction mod $N$. The product $Q_1\otimes \ket0$ takes the form
\begin{equation}
    Q_1\otimes \ket0=\sum_{i=0}^{N-1}\ket{i,0}\bra{i,i\oplus 1}.
\end{equation}
Now this square matrix can be rewritten as the product of a diagonal projector and a square permutation,
\begin{equation}\label{eq:Bpadpmr}
\begin{aligned}
    \sum_{i=0}^{N-1}\ket{i,0}\bra{i,i\oplus 1}&=\left[\mathcal{F}_1\otimes\ket{0}\bra{0}\right]\\&\quad\times\left[\sum_{i,j=0}^{N-1}\ket{i,j}\bra{i,i\oplus j\oplus1}\right].
\end{aligned}
\end{equation}
The first operator is diagonal, and the second is exactly the inverse of a quantum modular adder---i.e.\ a quantum subtractor---which we denote by $S_1$. Under periodic boundary conditions the diagonal projector reduces to $\mathds{1}\otimes\ket{0}\bra{0}$ (no boundary mask is needed); for notational uniformity we set it to $\mathcal{F}_1$, given by the explicit diagonal projectors in the open-boundary case. The analogous decomposition holds for $Q_{-1}\otimes \ket0$, where $\oplus1$ is simply replaced by $\ominus1$, and where we denote the corresponding adder and diagonal as $S_{-1}$ and $\mathcal{F}_{-1}$ respectively.

When generalizing the above PMR to $Q_{\pm1}^{(j,k)}$, an additional register shift is needed. We discuss and derive this generalization in Appendix~\ref{a:pmr}.

We define the permutation on the padded $k$-th subspace 
\begin{equation}\label{eq:Bpermutation}
S_{\pm1}^{(j,k,L)}:=\left[S_{\pm1}^{(j,k+1)}R^{(j+1,k+1)}\right]\otimes
\mathds{1}^{\otimes L-k-1},
\end{equation}
where $S_{\pm1}^{(j,k+1)}$ is the adder/subtractor acting between the $j$ and $k+1$-th registers, and $R^{(j+1,k+1)}$ denotes the cyclic left rotation of registers $j+1,...,k+1$, implementable with SWAPs. Note that the rotation adds $\mathcal{O}(L\log N)$ SWAP gates per application, which we will retain in our final resource estimates and absorb the $\log N$ factor into the soft-$\tilde{\mathcal{O}}$.

Let us also denote the diagonal on the padded subspace by 
\begin{equation}\label{eq:Bdiagonal}
    \mathcal{F}_{\pm1}^{(j,k,L)}:=\left[\mathcal{F}_{\pm1}^{(j,k)}\otimes\ket0\bra0\otimes\mathds1^{\otimes L-k-1}\right].
\end{equation}
The full PMR expansion of the padded matrix $\tilde{B}'$ is:
\begin{equation}
\begin{aligned}
    \tilde{B}'&=\frac{-1}{2a}\sum_{k=1}^{L-1}\sum_{j=1}^k(\Lambda_k\otimes \mathcal{F}_1^{(j,k,L)})(P_1^{(L)}\otimes S_1^{(j,k,L)})\\&\quad+\frac{1}{2a}\sum_{k=1}^{L-1}\sum_{j=1}^k(\Lambda_k\otimes \mathcal{F}_{-1}^{(j,k,L)})(P_1^{(L)}\otimes S_{-1}^{(j,k,L)}),
\end{aligned}
\end{equation}
where the $P_1^{(L)}$ is the $L\times L$ shift operator. The derivation of the above form can be found in Appendix~\ref{a:pmr}.

Before moving on, we recall that one of the underlying assumptions of the LCHS algorithm is that $G\succeq 0$. For the matrix we constructed for the Burgers' equation, it is not, however, guaranteed that the Hermitian part is positive semidefinite. This condition can be enforced by an overall shift of the spectrum, i.e., by adding a suitable multiple of the identity.
In Appendix~\ref{a3}, we prove that a shift of $L/a\sqrt2$ suffices, i.e., that the Hermitian part of the matrix $-(\tilde{A}'+\tilde{B}')+\frac{L}{a\sqrt2}\mathds1^{(L)}$ is positive semidefinite. Here $\mathds1^{(L)}=\mathds1_L\otimes\mathds1^{\otimes L}$ is the $LN^L\times LN^L$ identity. We note that this shift is entirely diagonal and does not alter the PMR expansion. Henceforth, we will include this shift in all our calculations. 

\subsection{LCHS-PMR implementation}\label{sec:lchspmr}

In the previous section, we showed that Burgers' equation can be reduced to a (truncated form of a) system of linear differential equations; in this subsection, we therefore first focus on constructing a quantum simulation algorithm that approximately solves Eq.~(\ref{eq:lchsODE}). We show that the PMR of a general matrix $X$ induces the PMR of every LCHS Hamiltonian $H+k_jG$ using the same family of permutations, with a common normalization $\tilde\Gamma\leq(1+K)\Gamma_X$. We then show that the PMR Hamiltonian simulation can be incorporated into LCHS using the available register $\ket{k_j}$ and using ancilla registers that encode phases and amplitudes of the PMR coefficients.

We start from the most general PMR expansion of $X$, allowing both Hermitian permutations (involutions) and non-Hermitian permutations (non-involutions):
\begin{equation}\label{xpmr}
    X=D_0+\sum_{i=1}^{\tilde{M}}D_i^{(1)}P_i+\sum_{i=1}^{\tilde{M}}D_i^{(2)}P_i^\dagger+\sum_{i=1}^{\tilde{\tilde{M}}}D_i^{(3)}\Pi_i\ ,
\end{equation}
where $P_i$ are non-Hermitian, $\Pi_i$  are Hermitian, and $D_i$ are diagonal masks. This form is convenient because it is invariant under the redefinitions $P_i\leftrightarrow P_i^\dagger$ and $D_i^{(1)}\leftrightarrow D_i^{(2)}$, and if the adjoint for any given $P_i$ is not a part of the PMR for $X$ it is encoded simply by setting the corresponding $D_i^{(2)}$ to zero.

Let $\Gamma_i^{(j)}\geq\norm{D_i^{(j)}}_{\max}$ be an upper bound on the elements of the diagonal masks, then
\begin{equation}
    \Gamma_X=\sum_{i=1}^{\tilde{M}}\Gamma_i^{(1)}+\sum_{i=1}^{\tilde{M}}\Gamma_i^{(2)}+\sum_{i=1}^{\tilde{\tilde{M}}}\Gamma_i^{(3)},
\end{equation}
where $\Gamma_X$ is denoted as the off-diagonal norm of $X$.

We show in Appendix~\ref{a:PMRLCHS} that the PMR expansion for $H+k_jG$ shares exactly the same permutation operators as the expansion for $X$, and that its off-diagonal norm is bounded by $\tilde\Gamma:=\Gamma_X(1+K)$. We adopt a common bound for all $j$ in order to ensure that the oblivious amplitude amplification works across all $j$, described in detail in Sec.~\ref{gatecost}. Note that asymptotically $\tilde\Gamma=\mathcal{O}(\Gamma_XK)$.

Define the short-time evolution $U^{(j)}:=U(\Delta t,k_j)$. From Eq.~(\ref{eq:Utilde}) the PMR approximation reads
\begin{equation}
    \tilde{U}^{(j)}=\sum_{q=0}^Q\frac{(\tilde\Gamma\Delta t)^q}{q!}\sum_{\mathbf i_q}\frac{\tilde\Gamma_{\mathbf i_q}}{(\tilde\Gamma)^q}\sum_{\mathbf k_q}\frac{1}{(K')^q}V_{(\mathbf i_q,\mathbf k_q)}^{(j)}\ ,
\end{equation}
with
\begin{equation}\label{eq:Vtilde}
    V_{(\mathbf i_q,\mathbf k_q)}^{(j)}=(-i)^qP_{\mathbf i_q}\sum_z\frac{d_{\mathbf i_q}^{(j)}(z)}{\tilde\Gamma_{\mathbf i_q}}e^{-i\delta \sum_{s=0}^q\alpha_sE_{z_s}^{(j)}}\ket{z}\bra{z}.
\end{equation}
The object $V_{(\mathbf i_q,\mathbf k_q)}^{(j)}$ is unitary if and only if the amplitude factors $d_{\mathbf i_q}^{(j)}(z)/\tilde\Gamma_{\mathbf i_q}$ have unit modulus for every $z$ on the relevant walk; in general $|d_{\mathbf i_q}^{(j)}(z)/\tilde\Gamma_{\mathbf i_q}|\leq 1$. The general case is handled by a phase-decomposition step that writes each amplitude as an average of two phases and absorbs the resulting factor of two into the LCU coefficients; this is described in detail in Appendix~\ref{a1} and does not alter the asymptotic complexity. For the remainder of this subsection we present the construction in the special case where each amplitude $d_{\mathbf i_q}^{(j)}(z)/\tilde\Gamma_{\mathbf i_q}$ is a pure phase, so that $V_{(\mathbf i_q,\mathbf k_q)}^{(j)}$ is itself unitary; the modifications required for the general case are summarized in the appendix.

The LCHS algorithm requires the controlled operation $\sum_{j=0}^{J-1}\ket{j}\bra{j}\otimes\tilde{U}^{(j)}$. The PMR structure---separating diagonal from off-diagonal evolutions---together with the availability of the $\ket{j}$ and $\ket{k_j}$ registers, makes this controlled operation natural to implement: the off-diagonal permutations $P_{\mathbf i_q}$ are $j$-independent, while all $j$-dependence enters through the diagonal evolution $\exp(-i\delta\alpha_sD_0^{(j)})$ and through the $j$-dependent state preparation discussed below. Because the LCHS quadrature uses Gaussian nodes inside each subinterval rather than a uniform grid on $[-K,K]$, the nodes $k_j$ are not equally spaced and we cannot in general write $\tilde U^{(j)}$ as the $j$th power of a single unitary. We instead implement the $k_j$-dependent diagonal evolution by reading $k_j$ coherently from the quadrature register $\ket{k_j}$ and synthesizing the phase $e^{-i\delta\alpha_s k_j D_0^{(G)}}$ via controlled arithmetic, following standard phase-synthesis circuits. Specifically, decomposing
\begin{equation}\label{eq:diagsplit}
    D_0^{(j)}=D_0^{(H)}+k_j\,D_0^{(G)},
\end{equation}
the diagonal phase factors as
\begin{equation}
    e^{-i\delta\alpha_sD_0^{(j)}}=e^{-i\delta\alpha_sD_0^{(H)}}\,e^{-i\delta\alpha_sk_jD_0^{(G)}},
\end{equation}
and the second factor is implemented by a phase-kickback circuit with control on $\ket{k_j}$. The cost of this controlled arithmetic depends on the precision $b$ used to encode $k_j$ and on the locality structure of $D_0^{(G)}$; we discuss this in Section~\ref{sec5}.

The off-diagonal coefficient ratios $d_{\mathbf i_q}^{(j)}(z)/\tilde\Gamma_{\mathbf i_q}$ also depend on $j$ in general, since
\begin{equation}\label{eq:dratio}
    \frac{d_i^{(j)}(z)}{\tilde\Gamma_i}=\frac{d_i^{(H)}(z)+k_j\,d_i^{(G)}(z)}{\tilde\Gamma_i}
\end{equation}
is a bounded complex amplitude whose magnitude and phase both depend on $k_j$. The procedure of Appendix~\ref{a1} handles this dependence via two ancilla registers $\ket{r_i^{(H)}}$ and $\ket{r_i^{(G)}}$, plus arithmetic with the available $\ket{k_j}$ register, to construct the appropriate phase decomposition coherently.

\section{Resource cost analysis}\label{sec5}

In this section we estimate the cost of the algorithm. We first treat the general construction of Sec.~\ref{sec:lchspmr}, collecting the approximations that enter it---the PMR truncation of each Hamiltonian simulation, the quadrature discretization of the LCHS integral, and the finite time segmentation---together with the parameters $Q$, $K$, $J$, and $r$ that control them, and assemble these into a gate count expressed through the off-diagonal norm $\Gamma_X$, the operator norm $\alpha_X$, and the number of PMR terms $M'$. We then specialize to the Carleman-linearized Burgers' system by evaluating each of these quantities for $X=-(\tilde A'+\tilde B')+\frac{L}{a\sqrt2}\mathds1^{(L)}$, and discuss the output model, the postselection overhead induced by the stabilizing shift, and the ancilla budget. Throughout, error is measured against the truncated, spatially discretized linear system of Eq.~(\ref{eq47}), not against the original nonlinear PDE; the spatial-discretization and Carleman-truncation errors are separate from our treatment.

\subsection{Resource cost of the general LCHS + PMR algorithm}

\subsubsection{Errors}

We now collect the errors of the algorithm and the parameters that control them. Let $\tilde{U}_j$ denote the long-time PMR approximation, and bound its error by $\epsilon_1$, i.e.\ $\norm{\tilde{U}_j-U(t,k_j)}\leq\epsilon_1$ for all $j$. Let the integral discretization error be bounded by $\epsilon_2$, i.e.\ $\norm{\sum_jc_jU(t,k_j)-e^{-Xt}}\leq\epsilon_2$. 

We note that in our analysis, we have not accounted for the error introduced by the spatial discretization and the truncation of the Carleman linearization procedure to finite order $L$. We compare our algorithmic approximation to the solution of the already linearized and truncated Eq.~(\ref{eq47}), not the solution to the nonlinear equation.

According to Ref.~\cite{Kalev_2025} and Lemmas~10 and 11 of Ref.~\cite{An_2025}, the following choices of $Q$, $K$, and $J$ suffice (under the conditions stated in those references):
\begin{equation}\label{eq:Qchoice}
    Q=\mathcal{O}\left(\frac{\log(\frac{t\tilde\Gamma}{\epsilon_1})}{\log\log(\frac{t\tilde\Gamma}{\epsilon_1})}\right)=\mathcal{O}\left(\frac{\log(\frac{tK\Gamma_X}{\epsilon_1})}{\log\log(\frac{tK\Gamma_X}{\epsilon_1})}\right),
\end{equation}
\begin{equation}\label{eq:Kchoice}
    K=\mathcal{O}\left(\left(\log\left(\frac{1}{\epsilon_2}\right)\right)^{1/\beta}\right),
\end{equation}
\begin{equation}\label{eq:Mchoice}
\begin{aligned}
    J&=\mathcal{O}\left(\norm{G}t\left(\log\left(\frac{1}{\epsilon_2}\right)\right)^{1+1/\beta}\right)\\&=\mathcal{O}\left(\alpha_X t\left(\log\left(\frac{1}{\epsilon_2}\right)\right)^{1+1/\beta}\right),
\end{aligned}
\end{equation}
where we have used $\tilde\Gamma=\mathcal{O}(\Gamma_XK)$ and $\alpha_X\geq\norm{X}\geq\norm{G}$.

The unnormalized LCHS error can be bounded by

\begin{equation}
\begin{aligned}
    &\bigg\|\sum_jc_j\tilde{U}_jy_0-e^{-Xt}y_0\bigg\|\\&=\bigg\|\sum_jc_j(\tilde{U}_j-U_j)y_0+\sum_jc_jU_jy_0-e^{-Xt}y_0\bigg\|\\&\leq\norm{y_0}\bigg(\bigg\|\sum_jc_j(\tilde{U}_j-U_j)\bigg\|+\bigg\|\sum_jc_jU_j-e^{-Xt}\bigg\|\bigg)\\&\leq\norm{y_0}\sum_j\norm{c_j(\tilde{U}_j-U_j)}+\norm{y_0}\epsilon_2\\&\leq\norm{y_0}\epsilon_1\sum_j|c_j|+\norm{y_0}\epsilon_2\\&=\norm{y_0}\norm{c}_1\epsilon_1+\norm{y_0}\epsilon_2.
\end{aligned}
\end{equation}
The induced error on the normalized quantum state then satisfies

\begin{equation}
\begin{aligned}
    \norm{\ket{v}-\ket{y(t)}}&\leq\frac{2}{\norm{y(t)}}\norm{v-y(t)}\\&\leq\frac{2\norm{y_0}\norm{c}_1}{\norm{y(t)}}\epsilon_1+\frac{2\norm{y_0}}{\norm{y(t)}}\epsilon_2,
\end{aligned}
\end{equation}
where $\ket{x}\equiv x/\norm{x}$. To bound the final state error by $\epsilon$, it is sufficient to take

\begin{equation}\label{eq:eps1choice}
        \epsilon_1=\frac{\norm{y(t)}}{4\norm{c}_1\norm{y_0}}\epsilon\ ,
\end{equation}
\begin{equation}\label{eq:eps2choice}
        \epsilon_2=\frac{\norm{y(t)}}{4\norm{y_0}}\epsilon\ .
\end{equation}
We retain $\norm{c}_1$ symbolically below and absorb the constant into the soft-$\tilde{\mathcal{O}}$ in our final estimates.

\subsubsection{Gate cost}\label{gatecost}

The gate cost of implementing the PMR approximation $\tilde{U}^{(j)}$ of $U(\Delta t,k_j):=e^{-i(H+k_jG)\Delta t}$ to error $\epsilon_1$, including the diagonal evolution, is given in Ref.~\cite{Kalev_2025} as
\begin{equation}
    \mathcal{O}\left(rQ\left(M'+\log\left(\frac{t\tilde\Gamma}{\epsilon_1}\right)+\tilde{L}\kappa_{\text{loc}}\right)\right),
\end{equation}
where $r$ is the number of time segments, $\tilde{L}$ is the number of terms in the diagonal Hamiltonian, and $\kappa_{\text{loc}}$ is the locality bound (the maximum number of tensored $Z$ operators per term). The choice $r=t\tilde\Gamma/\ln 2$ of Ref.~\cite{Kalev_2025} ensures that oblivious amplitude amplification can be used for all nodes $j$. The bound from Ref.~\cite{Kalev_2025} also fixes the divided-difference subdivision parameter $K'$ in terms of $\tilde\Gamma$, $t$, and $\epsilon_1$, so its contribution is already incorporated. 

The controlled-on-$j$ structure required by LCHS adds the cost of $k_j$-controlled arithmetic on the diagonal evolution of $D_0^{(G)}$. Encoding $k_j$ in a $b$-qubit register and synthesizing the phase $e^{-i\delta\alpha_sk_jD_0^{(G)}}$ by standard reversible-arithmetic circuits incurs a multiplicative overhead $\mathrm{poly}(b,\tilde L,\kappa_{\text{loc}})$ on the diagonal evolution, which we hide in the soft-$\tilde{\mathcal{O}}$ below. Thus the diagonal piece carries an extra factor $\tilde{\mathcal{O}}(\mathrm{poly}(b,\tilde L\kappa_{\text{loc}}))$ beyond the base PMR cost:
\begin{equation}
    \mathcal{O}\left(rQ\left(M'+\log\left(\frac{t\tilde\Gamma}{\epsilon_1}\right)+\tilde L \kappa_{\text{loc}}\cdot\mathrm{poly}(b)\right)\right).
\end{equation}
With $\tilde\Gamma=\mathcal{O}(\Gamma_XK)$, and including the amplitude amplification overhead and $C_{\mathrm{prep}}$ the state preparation cost for the LCU of Eq.~\eqref{eq:lcusum}, the overall gate cost of LCHS+PMR is
\begin{equation}
\begin{aligned}
    \mathcal{O}\Bigg(\Big(C_{\mathrm{prep}}+t&K\Gamma_XQ\big(M'+\log\big(\tfrac{tK\Gamma_X}{\epsilon_1}\big)\\&+\tilde L\kappa_{\text{loc}}\cdot\mathrm{poly}(b)\big)\Big)\times\frac{\norm{y_0}\norm{c}_1}{\norm{y(t)}}\Bigg).
\end{aligned}
\end{equation}
We provide a state preparation in Appendix~\ref{app:Cprep} such that $C_{\text{prep}}=\tilde{\mathcal{O}}(\sqrt K)$. Substituting Eqs.~(\ref{eq:Qchoice}--\ref{eq:Mchoice}) and Eqs.~(\ref{eq:eps1choice},\ref{eq:eps2choice}), and absorbing logarithmic factors, $C_{\mathrm{prep}}$ and $\norm{c}_1$ into the soft-$\tilde{\mathcal{O}}$ notation, this reduces to
\begin{equation}\label{eq:gencost}
\begin{aligned}    \tilde{\mathcal{O}}\left(t\Gamma_XM'\left(\log\left(\frac{1}{\epsilon}\right)\right)^{1+1/\beta}\frac{\norm{y_0}}{\norm{y(t)}}\right).
\end{aligned}
\end{equation}
The soft-$\tilde{\mathcal{O}}$ here hides polylogarithmic factors in $N$, $L$, $\alpha_X$, $1/\epsilon$, the quadrature precision $b$, and the reversible-arithmetic costs of modular adders, comparators, and phase synthesis. 

We emphasize that $\epsilon$ in this estimate is the error against the truncated lifted system Eq.~(\ref{eq47}), not against the original nonlinear PDE. The total error against Burgers' equation is bounded as
\begin{equation}
    \epsilon_{\mathrm{total}}\leq\epsilon_{\mathrm{spatial}}+\epsilon_{\mathrm{Carl}}+\epsilon\ ,
\end{equation}
where $\epsilon_{\mathrm{spatial}}$ is the spatial finite-difference truncation error and $\epsilon_{\mathrm{Carl}}$ is the Carleman truncation error at level $L$. Bounds on $\epsilon_{\mathrm{Carl}}$ under suitable assumptions on $\norm{\bu(t)}$, $t$, and $\nu$ are available in the literature~\cite{Liu_2021,Krovi_2023}; we do not reproduce them here.

\subsection{Application of the algorithm to Burgers' equation}\label{sec:burgersresources}

We now specialize Eq.~(\ref{eq:gencost}) to the Carleman-linearized Burgers' system, by counting the parameters $M'$, $\Gamma_X$, $\tilde L$, $\kappa_{\text{loc}}$, and $\alpha_X$ for $X=-(\tilde A'+\tilde B')+\frac{L}{a\sqrt2}\mathds1^{(L)}$. The negative sign outside $(\tilde{A}'+\tilde{B}')$ may be omitted as it will be absorbed into the diagonal masks, and is irrelevant for norm calculations.

The number of terms in the PMR expansion of $X$ must include adjoint terms as well, even when the corresponding diagonal coefficients vanish, since the construction in Sec.~\ref{sec3} requires it. The adjoint terms for $\tilde A'$ are already present, so only the extra adjoints for $\tilde B'$ contribute. The total count is $2L+4\frac{(L^2-L)}{2}=2L^2=\mathcal{O}(L^2)$. 

The off-diagonal norm $\Gamma_X$ similarly includes the adjoint terms, but since the diagonal coefficients of these adjoints vanish in the Burgers' case, they may be omitted from the norm. We obtain
\[
\Gamma_X=2L\frac{\nu}{a^2}+2\,\frac{L^2-L}{2}\frac{1}{2a}=\mathcal{O}\!\left(\tfrac{L\nu}{a^2}+\tfrac{L^2}{a}\right).
\]
For a periodic grid of total length $\zeta=Na$, this gives the full scaling
\begin{equation}\label{eq:GammaXfull}
    \Gamma_X=\mathcal{O}\!\left(\frac{\nu LN^2}{\zeta^2}+\frac{L^2N}{\zeta}\right).
\end{equation}
The diffusive and advective contributions cross at $\nu N/\zeta\sim 1$. In the diffusion-dominated regime $\nu N/\zeta\gg 1$, Eq.~(\ref{eq:GammaXfull}) simplifies to $\Gamma_X=\mathcal{O}(LN^2\nu/\zeta^2)$; in the advection-dominated regime $\nu N/\zeta\ll 1$, it simplifies to $\Gamma_X=\mathcal{O}(L^2N/\zeta)$. We retain both terms in what follows and quote final scalings under each regime where appropriate.

The diagonal piece of the PMR for $\tilde{A}'$ simplifies to
\begin{equation}
\begin{aligned}
    D_0&=\frac{-2\nu}{a^2}\sum_{k=1}^Lk(\Lambda_k\otimes\mathds1^{\otimes L})\\&=\frac{-2\nu}{a^2}\left(\sum_{k=1}^Lk\Lambda_k\right)\otimes\mathds1^{\otimes L}.
\end{aligned}
\end{equation}
Setting $L=2^l$, the operator in parentheses can be written using single-qubit operators as
\begin{equation}
\begin{aligned}
    (2^{l-1}+2^{-1})\mathds1^{\otimes l}-\sum_{i=0}^{l-1}2^{l-i-2}Z_i\ .
\end{aligned}
\end{equation}
Note that the additional eigenvalue shift can simply be included with the identity term above. Thus $\tilde{L}=\log L$ and $\kappa_{\text{loc}}=1$.

The operator norm for $X$ is bounded by: 
\begin{equation}
\begin{aligned}
    \norm{-(\tilde{A}'+\tilde{B}')+\frac{L}{a\sqrt2}\mathds1^{(L)}}&\leq \norm{\tilde{A}'+\tilde{B}'}+\frac{L}{a\sqrt2}\\&\leq L\left(\frac{4\nu}{a^2}+\frac{\sqrt2}{a}\right),
\end{aligned}
\end{equation}
so that $\alpha_X=\mathcal{O}(LN^2\nu/\zeta^2+LN/\zeta)$ in the diffusion-dominated and advection-dominated regimes respectively. The proof for the bound on $\norm{\tilde{A}'+\tilde{B}'}$ can be found in Appendix~\ref{a2}. The scaling of all parameters of the Burgers' equation is summarized in Table~\ref{tab:burgercost}.

\begin{table*}[t]
\centering
\renewcommand{\arraystretch}{2}
\setlength{\tabcolsep}{15pt}

\begin{tabular}{| c | c | c |}
\hline\hline

Quantity & Burgers value & Asymptotic scaling \\ [0.5ex]
\hline

$M'$
& $2L^2$
& $O(L^2)$ \\[1ex]
\hline

$\Gamma_X$
& $\dfrac{2\nu L}{a^2}+\dfrac{L^2-L}{2a}$
& $O\left(\dfrac{\nu LN^2}{\zeta^2}
+\dfrac{L^2N}{\zeta}\right)$ \\[1ex]
\hline

$\widetilde{\Gamma}$
& $(1+K)\Gamma_X$
& $O(K\Gamma_X)$ \\[0.5ex]
\hline

$\alpha_X$
& $L\left(\dfrac{4\nu}{a^2}+\dfrac{\sqrt{2}}{a}\right)$
& $O\left(L\left(\dfrac{\nu N^2}{\zeta^2}
+\dfrac{N}{\zeta}\right)\right)$ \\[1ex]
\hline

$\widetilde{L}$
& $\log L$
& $O(\log L)$ \\[0.5ex]
\hline

$\kappa_{\text{loc}}$
& $1$
& $O(1)$ \\[0.5ex]
\hline

$\sigma$
& $\dfrac{L}{a\sqrt{2}}$
& $O\left(\dfrac{LN}{\zeta}\right)$ \\[1ex]
\hline
\end{tabular}
\caption{\label{tab:burgercost}Resource parameters for the Carleman-linearized Burgers' equation.}
\end{table*}

Substituting into Eq.~(\ref{eq:gencost}), the final gate cost for Burgers' (in the diffusion-dominated regime) including the $\tilde{\mathcal O}(L)$ cost from the register rotations is:
\begin{equation}\label{eq:burgerscost}
\begin{aligned}
    \tilde{\mathcal{O}}\!\left(tL^4N^2\frac{\nu}{\zeta^2}\left(\log\!\left(\frac{1}{\epsilon}\right)\right)^{1+1/\beta}\frac{e^{Lt/a\sqrt2}\norm{\tilde\bu_0}}{\norm{\tilde\bu(t)}}\right),
\end{aligned}
\end{equation}
with the obvious modifications in the advection-dominated regime. We note that the value of the Carleman truncation $L$ depends on $\epsilon_{\text{Carl}}$ and the nonlinearity, and is typically chosen to be small. It is independent of the gridsize $N$. We discuss the additional factor $e^{Lt/a\sqrt2}$ in more detail in Sec.~\ref{sec:postsel}.

\subsubsection{Output model}\label{sec:output}

The algorithm prepares a normalized quantum state $\ket{\tilde\bu(t)}$ proportional to the Carleman-lifted, padded solution at time $t$, in the encoding of Eq.~(\ref{eq:utildestate}). The physical solution $\bu(t)$ is recovered, up to normalization, by projecting the outer label register onto $\ket{0}$ (the $k=1$ Carleman sector) and the padding registers onto $\ket{0}^{\otimes L-1}$. This projection succeeds with amplitude $\norm{\bu(t)}/\norm{\tilde\bu(t)}$, so preparing the normalized \emph{physical} state $\ket{\bu(t)}$ carries an additional amplitude-amplification factor of $\norm{\tilde\bu(t)}/\norm{\bu(t)}$ beyond Eq.~(\ref{eq:burgerscost}); the combined physical-state factor is thus $e^{Lt/a\sqrt2}\,\norm{\tilde\bu_0}/\norm{\bu(t)}$, while Eq.~(\ref{eq:burgerscost}) as written is the cost of preparing the lifted state. Reading out all $N$ amplitudes of $\bu(t)$ classically requires $\Omega(N)$ samples and removes the potential quantum advantage, which is therefore restricted to estimating global observables, inner products, or low-rank functionals of the solution---e.g., overlaps with target states, integrated moments, or smooth functionals computable via amplitude estimation.

\subsubsection{Postselection norm}\label{sec:postsel}

The factor $\norm{\tilde\bu_0}/\norm{\tilde\bu(t)}$ in Eq.~(\ref{eq:burgerscost}) reflects the postselection cost of recovering the normalized state $\ket{\tilde\bu(t)}$ from the LCHS construction; here $\norm{\tilde\bu(t)}$ denotes the norm of the \emph{unshifted} lifted solution, the effect of the shift being displayed separately through the exponential factor. This ratio is close to $\norm{\bu_0}/\norm{\bu(t)}$ only when the lifted-state norm is dominated by the $k=1$ sector. Writing $r=\norm{\bu}$, the weight of the $k=1$ sector is
\begin{equation}
    p_{k=1}=\frac{r^2}{\sum_{k=1}^Lr^{2k}}=\frac{1-r^2}{1-r^{2L}}\ ,
\end{equation}
which is $\Theta(1)$ for $r$ bounded away from $1$ but tends to $1/L$ as $r\to1^-$; domination is therefore not automatic for $r<1$. Since $p_{k=1}\geq1/L$ for all $r\leq1$, the associated amplitude overhead is at most $\sqrt L$ in this regime. For $r>1$ the higher Carleman blocks dominate and the relevant ratio remains $\norm{\tilde\bu_0}/\norm{\tilde\bu(t)}$, which may be substantially larger than $\norm{\bu_0}/\norm{\bu(t)}$.

The eigenvalue shift discussed earlier introduces an additional exponential factor in the postselection norm. According to the LCHS formula, an identity shift $\sigma\mathds1$ in the Hamiltonian simulations is equivalent to introducing the factor $e^{-\sigma t}$ into the propagator. This factor is not innocuous: it changes the norm of the unnormalized state $u(t)$. In the case of the Burgers' equation, the entire Carleman lifted matrix is shifted by the value $L/a\sqrt2$. Thus a factor of $e^{Lt/a\sqrt2}$ appears alongside $\norm{\tilde\bu_0}/\norm{\tilde\bu(t)}$. We show in Appendix~\ref{a4} a rescaling scheme based on Ref.~\cite{Costa_2025} that can significantly reduce the cost due to the postselection factor.

\subsubsection{Ancilla count}

The algorithm uses three sets of ancilla qubits. The PMR subroutine requires
\begin{equation}
\begin{aligned}
    &\mathcal{O}\!\left(
    M' + \log\!\left\lceil\frac{tK\Gamma_X}{\epsilon_1}\right\rceil
    \right)
    \\&=
    \mathcal{O}\!\left(
    L^2 + \log\!\left\lceil
    \frac{tKLN^2\nu}{\zeta^2\epsilon_1}
    \right\rceil
    \right)
\end{aligned}
\end{equation}
ancilla qubits in the diffusion-dominated regime, up to lower-order logarithmic factors. The LCU subroutine requires the quadrature-index register $\ket{j}$, of size $\mathcal{O}(\log\lceil J\rceil)$, together with a $b$-qubit quadrature register $\ket{k_j}$ encoding the node $k_j$ to precision $b$. The phase-decomposition step in Appendix~\ref{a1} contributes additional arithmetic registers of size $\mathcal{O}(\log(1/\epsilon_1))$ for storing the amplitudes $r_i^{(H)}$, $r_i^{(G)}$, and the derived phases $\theta_{i_s}^{(j)}$, $\phi_{i_s}^{(j)}$. Altogether, the algorithm thus acts on $L\log_2N$ system qubits and $\log_2L$ label qubits, together with the ancilla registers listed above.

We note that the number of LCU terms $J$ is affected by the identity shift through the norm bound $\alpha_X$, since $J=\mathcal{O}(\alpha_X t\log^{1+1/\beta}(1/\epsilon_2))$. For Burgers' equation the stabilizing shift does not change $\alpha_X$ asymptotically in either regime: the shift $\sigma=LN/(\zeta\sqrt2)$ is of the same order as the advective contribution $\mathcal{O}(LN/\zeta)$ to the unshifted norm bound, and is subdominant to the diffusive contribution $\mathcal{O}(LN^2\nu/\zeta^2)$ when the latter dominates. The quadrature cutoff $K$ is also affected due to the identity shift as indicated by Eq.~(\ref{eq:eps2choice}) and Eq.~(\ref{eq:Kchoice}). However, this change is polylogarithmic in the postselection factor, which we absorb into the soft-$\tilde{\mathcal{O}}$ notation since it already contains the full postselection factor.

\section{Extensions to more complex fluid equations}\label{sec6}

The Burgers' equation provides a useful test case because it contains the two structural ingredients that make nonlinear fluid equations challenging: a nonlinear advective term and a dissipative differential operator. From the perspective of the present algorithm, however, the important feature is not specific to Burgers' equation. The relevant structure is that, after spatial discretization, the nonlinear PDE becomes a finite-dimensional polynomial system of ordinary differential equations whose linear and nonlinear terms are built from local finite-difference stencil operations~\cite{LeVeque_2007}. Once this is true, the same sequence of steps used above can be repeated: spatial discretization, Carleman linearization, padding to a qubit-compatible block structure, PMR decomposition into diagonal masks and arithmetic permutations, and finally LCHS implementation of the resulting non-unitary propagator (cf.~Fig.~\ref{figpipeline}).

To make this statement more precise, consider a general semidiscretized fluid equation of the form
\begin{equation}
    \dot{u}
    =
    A u
    +
    \sum_{\ell=2}^{p} B^{(\ell)} u^{\otimes \ell}
    +
    b(t),
    \label{eq:general-polynomial-fluid-system}
\end{equation}
where \(u(t) \in \mathbb{C}^{mN}\) collects \(m\) fluid variables on \(N\) spatial grid points, \(A\) contains the linear differential operators, and \(B^{(\ell)}\) encodes the degree-\(\ell\) polynomial nonlinearities. The Burgers' equation studied above corresponds to the case \(m=1\), \(p=2\), with \(A\) being the discrete diffusion operator and \(B^{(2)}\) being the quadratic advection tensor. More complicated fluid models increase the number of fluid variables \(m\), the number of spatial dimensions, the stencil size, the number of polynomial couplings, or the polynomial degree \(p\), but they do not change the basic algorithmic structure.

For example, on a Cartesian grid in \(d\) spatial dimensions, first- and second-derivative stencils take the form
\begin{equation}
\begin{aligned}
    \partial_{\mu} u_a(x)
    \approx
    \sum_{s \in \mathcal{S}_{\mu}^{(1)}} c_{\mu,s}^{(1)}
    u_a(x+s),
    \\
    \Delta u_a(x)
    \approx
    \sum_{s \in \mathcal{S}^{(2)}} c_{s}^{(2)}
    u_a(x+s),
    \label{eq:general-stencil}
\end{aligned}
\end{equation}
where \(a\) labels the fluid component and \(x=(x_1,\dots,x_d)\) is a \(d\)-dimensional grid point, each component being an integer index taken modulo the number of grid points in that direction under periodic boundary conditions. The offsets \(s\in\mathbb{Z}^d\) are the lattice displacements of the stencil, and \(x+s\) is understood component-wise modulo the grid. The set \(\mathcal{S}_{\mu}^{(1)}\) collects the offsets of the first-derivative stencil along direction \(\mu\); for the centered scheme used here, \(\mathcal{S}_{\mu}^{(1)}=\{\pm e_\mu\}\) with coefficients \(c_{\mu,\pm e_\mu}^{(1)}=\pm 1/2a\). Likewise \(\mathcal{S}^{(2)}\) collects the offsets of the Laplacian stencil \(\Delta=\sum_\mu \partial_\mu^2\); in the simplest case \(s=0\) carries weight \(-2d/a^2\) and \(s=\pm e_\mu\) carries weight \(1/a^2\). Because the position vector \(x\) already carries all \(d\) spatial components, no additional directional index on \(x\) or \(s\) is needed; the direction index appears only on \(\partial_\mu\), \(\mathcal{S}_{\mu}^{(1)}\), and \(c_{\mu,s}^{(1)}\). Each shift \(x \mapsto x+s\) is represented by a multidimensional modular shift operator. Thus the linear part of the discretized equation admits a PMR expansion of the same type as in the Burgers' construction,
\begin{equation}
    A
    =
    D_0
    +
    \sum_{\alpha} D_{\alpha} P_{\alpha},
    \label{eq:general-linear-pmr}
\end{equation}
where \(P_{\alpha}\) are either spatial shifts \(\sum_x\ket{x\oplus\alpha}\bra{x}\), component swaps, or combinations thereof, and \(D_{\alpha}\) are diagonal coefficient masks. Concretely, each stencil offset \(s\) maps to a shift permutation and each finite-difference coefficient maps to an entry of the corresponding diagonal mask. Boundary conditions are handled by the same mechanism as above: periodic boundary conditions give pure modular permutations, and for constant-coefficient stencils the accompanying masks \(D_\alpha\) reduce to multiples of the identity. Spatially varying coefficients, such as a position-dependent advection velocity, or a passage to spectral variables produce nontrivial diagonal masks even under periodic boundaries, while open, inflow, outflow, or mixed boundary conditions introduce additional diagonal masks that restrict the action of the corresponding permutation near the boundary.

The nonlinear terms have an analogous structure. A typical degree-\(\ell\) local nonlinear stencil term has the form
\begin{equation}
    u_{a_1}(x+s_1) u_{a_2}(x+s_2) \cdots u_{a_{\ell}}(x+s_{\ell}),
    \label{eq:local-monomial}
\end{equation}
possibly multiplied by a constant, a spatially dependent coefficient, or a component-dependent tensor. The corresponding rectangular map from \(u^{\otimes \ell}\) to \(u\) can be written as
\begin{equation}
\begin{aligned}
    &Q^{a; a_1,\ldots,a_{\ell}}_{s_1,\ldots,s_{\ell}}
    \\&=
    \sum_x
    |a,x\rangle
    \langle a_1,x+s_1;\,a_2,x+s_2;\,\ldots;\,a_{\ell},x+s_{\ell} | .
    \label{eq:general-Q-map}
\end{aligned}
\end{equation}
This is the direct analogue of the \(Q_{\pm 1}\) maps used for the Burgers' nonlinearity. After padding the rectangular map to the common Carleman block dimension, it decomposes into a diagonal mask times a reversible arithmetic permutation,
\begin{equation}
    Q^{a; a_1,\ldots,a_{\ell}}_{s_1,\ldots,s_{\ell}}
    \otimes \ket{0}
    =
    D^{a; a_1,\ldots,a_{\ell}}_{s_1,\ldots,s_{\ell}}
    P^{a; a_1,\ldots,a_{\ell}}_{s_1,\ldots,s_{\ell}},
    \label{eq:general-Q-pmr}
\end{equation}
where \(P^{a; a_1,\ldots,a_{\ell}}_{s_1,\ldots,s_{\ell}}\) is implemented by modular additions, modular subtractions, component relabelings, zero-padding checks, and register shifts, while \(D^{a; a_1,\ldots,a_{\ell}}_{s_1,\ldots,s_{\ell}}\) enforces boundary conditions and coefficient factors. Thus the same PMR primitives that appear in the Burgers' example---shifts, adders, subtractors, and diagonal masks---also implement the local polynomial couplings of more general fluid systems.

\begin{figure*}[htb]
\centering
	\includegraphics[width=1.0\linewidth]{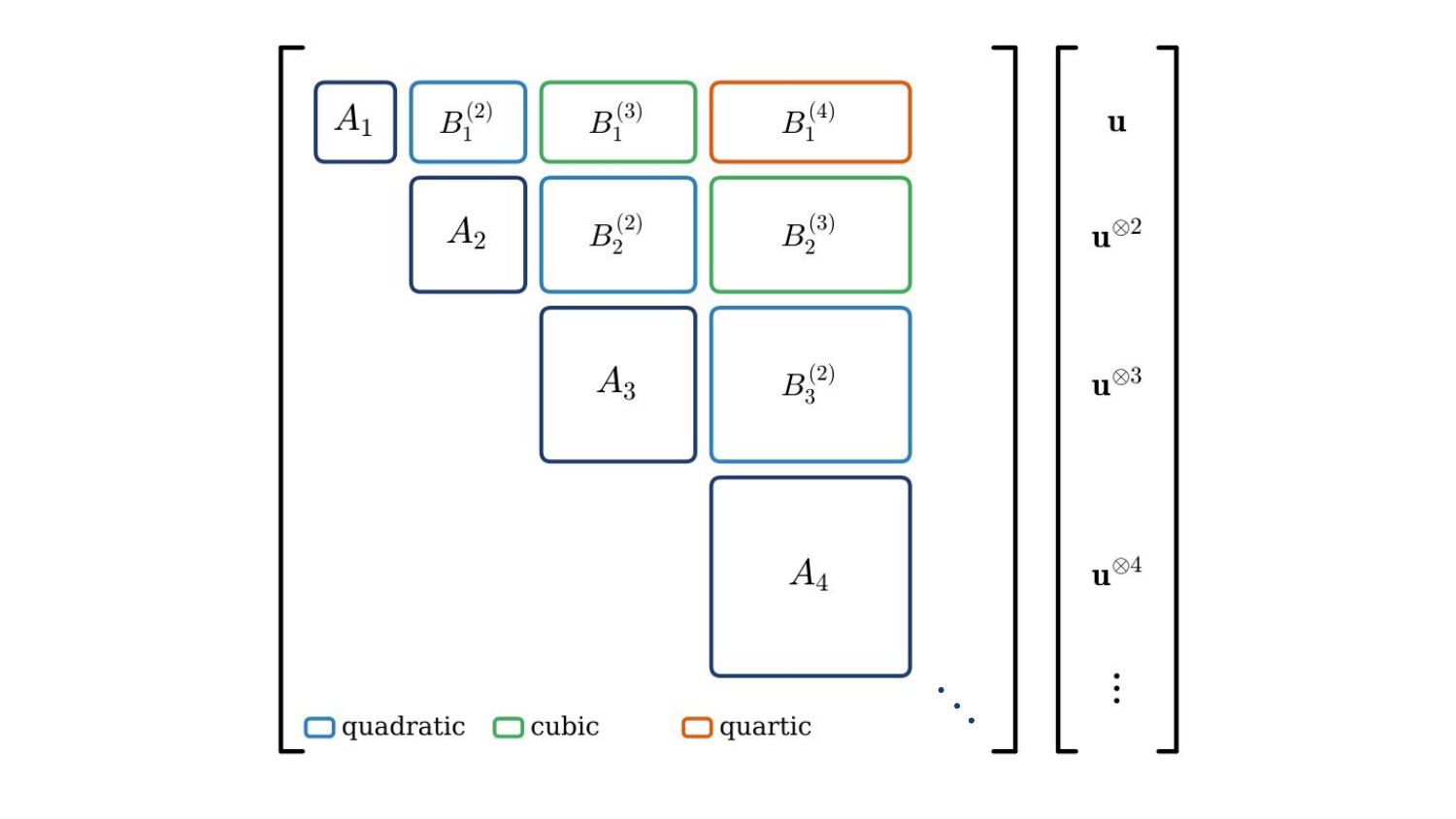}
	\caption{Structure of the matrix $(\tilde{A} + \tilde{B})$ for general higher order nonlinearities. The diagonal blocks $A_k$ implement the linear part of the dynamics within the $\bu^{\otimes k}$ sector, while the off-diagonal blocks $B_k^{(l)}$ couple $\bu^{\otimes k}$ to $\bu^{\otimes(k+l-1)}$. The quadratic coupling blocks $B_k$ of Fig.~\ref{figblock} are the degree-two special case $B_k^{(2)}$.}
\label{figblock2}
\end{figure*}

The Carleman hierarchy also generalizes directly.  For Eq.~\eqref{eq:general-polynomial-fluid-system}, one obtains
\begin{equation}
    \frac{d}{dt} u^{\otimes k}
    =
    \sum_{r=1}^{k}
    u^{\otimes (r-1)}
    \otimes \dot{u}
    \otimes
    u^{\otimes (k-r)} .
    \label{eq:general-carleman-start}
\end{equation}
Substituting Eq.~\eqref{eq:general-polynomial-fluid-system} gives
\begin{equation}
\begin{aligned}
    \frac{d}{dt} u^{\otimes k}
    =
    A_k u^{\otimes k}
    &+
    \sum_{\ell=2}^{p}
    B^{(\ell)}_k u^{\otimes(k+\ell-1)}
    \\&+
    \text{inhomogeneous terms},
    \label{eq:general-carleman}
\end{aligned}
\end{equation}
where
\begin{equation}
    A_k
    =
    \sum_{r=1}^{k}
    1^{\otimes(r-1)}
    \otimes A
    \otimes
    1^{\otimes(k-r)}
    \label{eq:general-Ak}
\end{equation}
and
\begin{equation}
    B^{(\ell)}_k
    =
    \sum_{r=1}^{k}
    1^{\otimes(r-1)}
    \otimes B^{(\ell)}
    \otimes
    1^{\otimes(k-r)} .
    \label{eq:general-Bk}
\end{equation}
The truncated lifted system therefore has a block structure similar to the Burgers' case, except that the block matrix is no longer only first-superdiagonal: a degree-\(p\) nonlinearity gives it upper block bandwidth \(p-1\), with the \((\ell-1)\)-st superdiagonal populated by the degree-\(\ell\) couplings \(B^{(\ell)}_k\). A degree-\(\ell\) nonlinearity couples the \(k\)-th Carleman sector to the \((k+\ell-1)\)-st sector. Consequently, quadratic nonlinearities produce nearest-neighbor couplings in Carleman level, cubic nonlinearities produce next-nearest-neighbor couplings, and so on as shown in Fig.~\ref{figblock2}. The inhomogeneous term \(b(t)\) in Eq.~\eqref{eq:general-polynomial-fluid-system} likewise generates a first-subdiagonal coupling, from sector \(k\) to sector \(k-1\); this is omitted from Fig.~\ref{figblock2}, which depicts the homogeneous part. Padding all retained sectors to a common dimension again produces a square linear system suitable for LCHS and PMR-based Hamiltonian simulation.

This observation gives a systematic route from Burgers' equation to richer fluid models. A first extension is the multidimensional scalar Burgers' equation,
\begin{equation}
    \partial_t u
    =
    \nu \Delta u
    -
    \sum_{\mu=1}^{d} u \, \partial_{\mu} u ,
    \label{eq:multidimensional-scalar-burgers}
\end{equation}
or the vector Burgers' equation,
\begin{equation}
    \partial_t u_a
    =
    \nu \Delta u_a
    -
    \sum_{\mu=1}^{d} u_{\mu} \partial_{\mu} u_a .
    \label{eq:vector-burgers}
\end{equation}
The diffusion term contributes multidimensional shifts \(P_{\pm e_\mu}\), while the advective term contributes quadratic maps of the form in Eq.~\eqref{eq:general-Q-map}. The only new ingredients relative to the one-dimensional scalar Burgers' equation are component labels and additional spatial shift directions. The PMR implementation is therefore obtained by replacing one-dimensional cyclic shifts and adders by their multidimensional and component-controlled analogues.

A second extension is to advection--diffusion--reaction systems~\cite{Budinski_2021},
\begin{equation}
    \partial_t c_a
    +
    \sum_{\mu=1}^{d} v_{\mu}(x,t) \partial_{\mu} c_a
    =
    D_a \Delta c_a
    +
    R_a(c_1,\ldots,c_m),
    \label{eq:reaction-advection-diffusion}
\end{equation}
where the reactions \(R_a\) are polynomial, or have first been approximated by polynomial expansions over the relevant dynamical range. If \(v\) is prescribed, the advection and diffusion terms are linear stencil operators and the reaction terms determine the polynomial degree of the Carleman hierarchy. If \(v\) is itself dynamical and coupled to the concentrations, the combined vector \(u=(v,c)\) still fits Eq.~\eqref{eq:general-polynomial-fluid-system}, provided the coupling terms are polynomial after discretization. In either case, local reactions become diagonal or low-degree Carleman couplings, while transport terms become shift-and-mask PMR terms.

A third important example is the incompressible Navier--Stokes equation~\cite{Gaitan_2020, Sanavio_2024},
\begin{equation}
    \partial_t \mathbf{u}
    +
    (\mathbf{u}\cdot\nabla)\mathbf{u}
    =
    -\nabla p
    +
    \nu \Delta \mathbf{u}
    +
    \mathbf{f},
    \qquad
    \nabla\cdot \mathbf{u}=0 .
    \label{eq:incompressible-ns}
\end{equation}
The nonlinear advection term is again quadratic and can be treated by the same Carleman mechanism as Burgers' equation. The additional issue is the pressure, which enforces the incompressibility constraint. One standard formulation removes the pressure by applying the Leray projection \(\Pi\) onto divergence-free vector fields~\cite{temam2001navier}:
\begin{equation}
    \partial_t \mathbf{u}
    =
    \nu \Delta \mathbf{u}
    -
    \Pi\big[(\mathbf{u}\cdot\nabla)\mathbf{u}\big]
    +
    \Pi \mathbf{f}.
    \label{eq:projected-ns}
\end{equation}
In a Fourier discretization, the diffusion operator and the incompressibility projector are diagonal in momentum space, while the quadratic advection term becomes a convolution constrained by momentum conservation. Such convolution terms are naturally implemented by modular adders over momentum registers, with diagonal masks enforcing component tensors and projection factors. Thus, in spectral variables, the Navier--Stokes extension replaces the real-space shift-and-mask structure of Burgers' equation by a momentum-space diagonal-and-adder structure. This remains compatible with PMR: the permutations are arithmetic maps enforcing momentum addition, and the coefficients are diagonal masks depending on the Fourier modes and vector components.

Compressible and shallow-water models~\cite{vreugdenhil1994numerical} provide another class of extensions. In primitive variables, a simplified shallow-water system may be written schematically as
\begin{align}
    \partial_t h
    &=
    -\nabla\cdot(h\mathbf{u}), \\
    \partial_t \mathbf{u}
    &=
    -(\mathbf{u}\cdot\nabla)\mathbf{u}
    -
    g\nabla h
    +
    \nu \Delta \mathbf{u},
    \label{eq:shallow-water}
\end{align}
which contains quadratic transport terms and linear gravity-wave and diffusion terms. After discretization, the \(h u_\mu\) and \(u_\mu \partial_\mu u_a\) terms are handled by the same quadratic Carleman blocks as above. More general compressible Euler or compressible Navier--Stokes systems can also be brought into this framework whenever the chosen variables and closure relations yield polynomial nonlinearities, or after non-polynomial constitutive relations, such as equations of state or transport coefficients, are approximated by polynomial expansions over the domain of interest.  In that case, higher polynomial degree increases the bandwidth of the Carleman block matrix but does not otherwise change the LCHS--PMR construction.

The same comments apply to magnetohydrodynamic-type systems~\cite{drake2006high} with polynomial couplings. For instance, the induction equation contains terms of the form
\begin{equation}
    \partial_t \mathbf{B}
    =
    \nabla \times (\mathbf{u}\times \mathbf{B})
    +
    \eta \Delta \mathbf{B},
    \label{eq:mhd-induction}
\end{equation}
which are quadratic in the velocity and magnetic fields, together with linear derivative operators. The solenoidal constraints on \(\mathbf{u}\) or \(\mathbf{B}\) can be treated using projection operators, analogously to the incompressible Navier--Stokes case, while the quadratic couplings again produce Carleman blocks expressible in terms of arithmetic permutations and diagonal coefficient masks.

These examples illustrate the general principle. The PMR--LCHS construction is not tied to the scalar one-dimensional Burgers' equation. The Burgers' equation is simply the smallest nontrivial instance of a broader pattern: spatial discretization turns a local polynomial PDE into a polynomial semidiscrete ODE, Carleman linearization lifts it to a finite linear system, the lifted generator is decomposed in the PMR representation, and the resulting propagator is implemented through LCHS. 

The price of moving up this hierarchy is predictable: more fields increase the component-register size, higher spatial dimension increases the number of stencil shifts and the spatial-register size, higher-order discretizations increase the number of shift terms, and higher-degree nonlinearities increase the Carleman coupling bandwidth. The underlying quantum primitives, however, remain the same: reversible arithmetic operations implementing shifts, additions, subtractions, and component swaps, together with diagonal coefficient masks. Therefore, whenever the spatially discretized fluid equation has polynomial nonlinearities and efficiently computable local or spectral coupling rules, the methodology developed here gives a direct recipe for constructing the corresponding PMR decomposition and for using it inside the LCHS framework.

One caveat applies uniformly across these extensions. The construction above establishes only that each model admits a PMR decomposition into arithmetic permutations and diagonal masks; it does not by itself guarantee the positive semidefiniteness condition \(G\succeq 0\) that LCHS imposes on the Hermitian part of the generator, possibly after a stabilizing shift. For the dissipation-dominated Burgers' system this condition is secured by the shift analysis of Appendix~\ref{a3}; there the shift leaves $\alpha_X$ and the polynomial PMR prefactor unchanged, but it does introduce the exponential postselection factor displayed explicitly in Eq.~(\ref{eq:burgerscost}), mitigated by the rescaling of Appendix~\ref{a4}. For advection-dominated regimes, the incompressible Navier--Stokes advection term, or magnetohydrodynamic couplings, positivity must be checked case by case, and any required shift feeds back into the amplitude-amplification cost \(\norm{u_0}/\norm{u(t)}\) in the same way it does for Burgers'. Establishing positivity, together with the accompanying Carleman convergence bounds, for these richer systems is part of what we leave to future work.

\section{Summary and conclusions}\label{sec7}

We have presented a PMR-based instantiation of the LCHS subroutine for the finite-dimensional linear systems obtained by Carleman linearization of polynomial nonlinear ODEs, with the spatially discretized viscous Burgers' equation as a worked example. The construction proceeds in three stages: a finite-dimensional Carleman linearization of the discretized nonlinear system into a linear ODE; a PMR decomposition of the padded block matrix that describes that linear system; and an LCHS-based implementation of the resulting non-unitary propagator, in which the family of Hamiltonian simulations required by LCHS is supplied by the PMR algorithm. Subject to the assumptions required by LCHS---most importantly positive semidefiniteness of the Hermitian part $G$ of the linear generator, possibly after a stabilizing shift---and conditional on suitable spatial-discretization and Carleman-truncation error bounds, the algorithm prepares, after amplitude amplification, a normalized quantum state proportional to the time-evolved solution of the truncated lifted system.

The PMR step is what ties the construction together. It represents the otherwise unwieldy matrix $\tilde A'+\tilde B'$ in terms of standard arithmetic primitives---spatial shift operators $P_{\pm 1}^{(L)}$, quantum modular adders and subtractors $S_{\pm 1}$, and diagonal masks---that are well within the toolbox of fault-tolerant quantum computing. PMR also integrates naturally with the LCHS controlled-block structure: the dependence on the LCU index $\ket{j}$ (equivalently the quadrature register $\ket{k_j}$) enters only through the diagonal evolution and through state-preparation rotations, both of which are handled by standard reversible-arithmetic circuits with at most polylogarithmic overhead.

The framework extends naturally to other polynomial systems whose Carleman blocks admit efficient PMR decompositions into arithmetic permutations and diagonal masks. Polynomial nonlinearities of higher order can be accommodated by extending the Carleman hierarchy and introducing additional shift-and-mask building blocks in the PMR expansion, with the same overall structure carrying through. Extensions to higher spatial dimensions, sharper Carleman convergence analyses (see, e.g., Refs.~\cite{Liu_2021, GonzalezConde_2025} for convergence conditions tying truncation error to the physical parameters of the flow), and applications to other nonlinear PDEs of physical interest are natural directions for future work.

\begin{acknowledgments}
We thank Ivan Bermejo-Moreno for many helpful discussions. Research was sponsored by the U.S. Army Research Laboratory and was accomplished under Cooperative Agreement Number
W911NF2410242. IH acknowledges
support by the Office of Advanced Scientific Computing Research of the
U.S. Department of Energy under Contract No.~DE-SC0024389.
AI-based language tools were used to assist with stylistic and
grammatical editing of the manuscript. No AI tools were
used to generate or modify the underlying mathematical results,
algorithms, proofs, or scientific conclusions.
\end{acknowledgments}

\bibliography{refs}

\appendix

\section{PMR for $H+k_jG$ and proof for the off-diagonal norm bound}\label{a:PMRLCHS}

In this appendix, we show that the PMR expansion for $H$ and $G$ shares exactly the same permutation operators as $X$ and the off-diagonal norm for both $H$ and $G$ is bounded by $\Gamma_X$. Consequently, the permutation operators for $H+k_jG$ are also exactly the same, and a common off-diagonal norm bound $\Gamma_X(1+K)$ can be fixed for all $j$.

We begin by taking the adjoint of Eq.~(\ref{xpmr}) and reordering, we have

\begin{equation}
\begin{aligned}
    X^\dagger&=D_0^\dagger+\sum_{i=1}^{\tilde{M}}P_i^\dagger {D_i^{(1)}}^\dagger+\sum_{i=1}^{\tilde{M}}P_i{D_i^{(2)}}^\dagger+\sum_{i=1}^{\tilde{\tilde{M}}}\Pi_i{D_i^{(3)}}^\dagger\\&=D_0^\dagger+\sum_{i=1}^{\tilde{M}}{D_i^{(1)}}'P_i^\dagger+\sum_{i=1}^{\tilde{M}}{D_i^{(2)}}'P_i+\sum_{i=1}^{\tilde{\tilde{M}}}{D_i^{(3)}}'\Pi_i,
\end{aligned}
\end{equation}
with ${D_i^{(1)}}'=P_i^\dagger{D_i^{(1)}}^\dagger P_i$, ${D_i^{(2)}}'=P_i{D_i^{(2)}}^\dagger P_i^\dagger$, and ${D_i^{(3)}}'=\Pi_i{D_i^{(3)}}^\dagger \Pi_i$. The diagonal elements of $D_i$ and $D_i'$ are the same up to a permutation and up to conjugations; in particular, $\norm{D_i^{(j)}}_{\max}=\norm{{D_i^{(j)}}'}_{\max}$ and
$\Gamma_X=\Gamma_{X^\dagger}$.

The Hermitian matrices $G$ and $H$ inherit a PMR expansion with the same set of permutations:

\begin{equation}
\begin{aligned}
    G:&=\frac{X+X^\dagger}{2}\\&=\frac{D_0+D_0^\dagger}{2}+\sum_{i=1}^{\tilde M}\frac{D_i^{(1)}+{D_i^{(2)}}'}{2}P_i\\&\quad\ +\sum_{i=1}^{\tilde M}\frac{D_i^{(2)}+{D_i^{(1)}}'}{2}P_i^\dagger+\sum_{i=1}^{\tilde{\tilde{M}}}\frac{D_i^{(3)}+{D_i^{(3)}}'}{2}\Pi_i,
\end{aligned}
\end{equation}

\begin{equation}
\begin{aligned}
    H:&=\frac{X-X^\dagger}{2i}\\&=\frac{D_0-D_0^\dagger}{2i}+\sum_{i=1}^{\tilde M}\frac{D_i^{(1)}-{D_i^{(2)}}'}{2i}P_i\\&\quad\ +\sum_{i=1}^{\tilde M}\frac{D_i^{(2)}-{D_i^{(1)}}'}{2i}P_i^\dagger+\sum_{i=1}^{\tilde{\tilde{M}}}\frac{D_i^{(3)}-{D_i^{(3)}}'}{2i}\Pi_i,
\end{aligned}
\end{equation}
Three observations are in order. First, although the expansions for $G$ and $H$ involve the same set of permutations as $X$, individual diagonal coefficients may vanish. Second, by the triangle inequality the diagonal coefficients of $G$ and $H$ obey identical \emph{upper bounds}:
\begin{equation}
    \norm{D_i^{(1)}\pm{D_i^{(2)}}'}\leq\Gamma_{i}^{(1)}+\Gamma_i^{(2)},
\end{equation}
\begin{equation}
    \norm{D_i^{(2)}\pm{D_i^{(1)}}'}\leq\Gamma_{i}^{(1)}+\Gamma_i^{(2)},
\end{equation}
\begin{equation}
    \norm{D_i^{(3)}\pm{D_i^{(3)}}'}\leq\Gamma_i^{(3)}.
\end{equation}
Third, summing these bounds, the off-diagonal norms of $G$ and $H$ are both bounded by $\Gamma_X$.

It is convenient to relabel the combined index set and write the PMR expansions of $G$ and $H$ in compact form:

\begin{equation}
    G:=D_0^{(G)}+\sum_{i=1}^{M'}D_i^{(G)}P_i'\ ,
\end{equation}
\begin{equation}
    H:=D_0^{(H)}+\sum_{i=1}^{M'}D_i^{(H)}P_i'\ ,
\end{equation}
where $D_0^{(G)}:=\frac{D_0+D_0^\dagger}{2}$, $D_0^{(H)}:=\frac{D_0-D_0^\dagger}{2i}$, $M'=2\tilde{M}+\tilde{\tilde{M}}$, and we choose the common upper bounds $\Gamma_i^{(G)}=\Gamma_i^{(H)}$ for each $i$, so that
\begin{equation}
    \Gamma^{(G)}=\Gamma^{(H)}=\Gamma_X.
\end{equation}

The PMR expansion of $H+k_jG$ then follows immediately:

\begin{equation}\label{eq:HplkL}
\begin{aligned}
    H+k_jG&=\left(D_0^{(H)}+k_jD_0^{(G)}\right)\\&\quad+\sum_{i=1}^{M'}\left(D_i^{(H)}+k_jD_i^{(G)}\right)P_i'\\&:=D_0^{(j)}+\sum_{i=1}^{M'}D_i^{(j)}P_i'.
\end{aligned}
\end{equation}
From this expansion, a common upper bound on the off-diagonal norm of $H+k_jG$ is $\tilde\Gamma=\Gamma_X(1+K)$, obtained by the triangle inequality $|D_i^{(H)}(z)+k_jD_i^{(G)}(z)|\leq\Gamma_i^{(H)}+|k_j|\Gamma_i^{(G)}\leq(1+K)\Gamma_i^{(H)}$.

\section{General case: $|d_i^{(j)}(z)/\tilde\Gamma_i|\leq1$}\label{a1}
This appendix follows the discussion in Sec.~3.5 of Ref.~\cite{Kalev_2025} and Appendix~D of Ref.~\cite{Kalev_2021}, adapted to the LCHS combination $H+k_jG$.

In general, the bound $\tilde\Gamma_i\geq\max_z|d_i^{(j)}(z)|$ guarantees
\begin{equation}
    \left|\frac{d_{i_s}^{(j)}(z)}{\tilde\Gamma_{i_s}}\right|\leq1\quad\forall\, s,z.
\end{equation}
We may therefore write the bounded amplitude as an average of two phases
\begin{equation}\label{eq:phasedecomp}
    \frac{d_{i_s}^{(j)}(z)}{\tilde\Gamma_{i_s}}=\frac{1}{2}\left(e^{i(\theta_{i_s}^{(j)}(z)+\phi_{i_s}^{(j)}(z))}+e^{i(\theta_{i_s}^{(j)}(z)-\phi_{i_s}^{(j)}(z))}\right)
\end{equation}
for suitable real $\theta_{i_s}^{(j)}(z),\phi_{i_s}^{(j)}(z)$, and modify the relevant unitary as
\begin{equation} 
    U_{(i_s,\mathbf{k}_q,\pm)}^{(j)}=-i\,\Phi _s^{(\pm)}\,e^{-i\delta\alpha_sD_0^{(H)}}\,e^{-i\delta\alpha_sk_jD_0^{(G)}}\,P_{i_s},
\end{equation}
where $\Phi _s^{(\pm)}=\sum_ze^{i(\theta_{i_s}^{(j)}(z)\pm\phi_{i_s}^{(j)}(z))}\ket{z}\bra{z}$ are diagonal phase unitaries. Following the discussion in Ref.~\cite{Kalev_2021}, $\Phi_s^{(\pm)}$ can be implemented by a phase-kickback circuit acting on registers $\ket{j}\ket{k_j}\ket{i_s}\ket{\theta_{i_s}^{(j)}}\ket{\phi_{i_s}^{(j)}}\ket{x}$, with $x\in\{0,1\}$ selecting the sign.

To construct the registers $\ket{\theta_{i_s}^{(j)}}$ and $\ket{\phi_{i_s}^{(j)}}$ we proceed as follows. With our choice $\tilde\Gamma_i=(1+K)\Gamma_i^{(H)}$ (and $\Gamma_i^{(H)}=\Gamma_i^{(G)}$), the amplitude ratio decomposes as
\begin{equation}\label{eq:rij}
\begin{aligned}
    r_{i_s}^{(j)}(z)\;:=\;\frac{d_{i_s}^{(j)}(z)}{\tilde\Gamma_{i_s}}
    \;&=\;\frac{d_{i_s}^{(H)}(z)+k_j\,d_{i_s}^{(G)}(z)}{(1+K)\Gamma_{i_s}^{(H)}}
    \;\\&=\;\frac{r_{i_s}^{(H)}(z)+k_j\,r_{i_s}^{(G)}(z)}{1+K},
\end{aligned}
\end{equation}
where $r_{i_s}^{(H)}(z):=d_{i_s}^{(H)}(z)/\Gamma_{i_s}^{(H)}$ and $r_{i_s}^{(G)}(z):=d_{i_s}^{(G)}(z)/\Gamma_{i_s}^{(G)}$ are normalized matrix elements of $D_{i_s}^{(H)}$ and $D_{i_s}^{(G)}$ respectively, with $|r_{i_s}^{(H)}(z)|,|r_{i_s}^{(G)}(z)|\leq 1$. Each is computable at the cost of evaluating one off-diagonal matrix element of $H$ or $G$~\cite{Kalev_2021}, so the registers $\ket{r_{i_s}^{(H)}}$ and $\ket{r_{i_s}^{(G)}}$ can be prepared by standard oracle access. Combining them with the available $\ket{k_j}$ register and the denominator $1+K$ via reversible arithmetic produces the register $\ket{r_{i_s}^{(j)}}$. We then follow the $U_{\mathrm{decomp}}$ procedure from Ref.~\cite{Kalev_2021} to obtain $\ket{\theta_{i_s}^{(j)}+(-1)^x\phi_{i_s}^{(j)}}$.

Additionally, the LCU state preparation must be modified to include $Q$ additional qubits each prepared in $\ket{+}$ to account for the $q$ extra factors of $1/2$ generated by Eq.~(\ref{eq:phasedecomp}). We note that these modifications do not alter the asymptotic complexity of the algorithm for the periodic boundary case; they contribute additional polylogarithmic factors that are absorbed into the soft-$\tilde{\mathcal{O}}$ in Sec.~\ref{sec5}. However for general boundary conditions the additional cost will need to be analyzed more carefully.

\section{Derivation of the Carleman linearized Burgers' equation}\label{a:carleman}

In this appendix, we derive the general form of the discretized Eq.~\eqref{eq:discreteburger}.

Extending the equation to another tensor power, we can see
\bea
\frac{\rmd \bu^{\otimes 2}}{\rmd t} &=& \bu \otimes\dot{\bu} + \dot{\bu} \otimes \bu
\nonumber\\&=& \bu \otimes \left(A \bu  + B \bu^{\otimes 2} \right) + \left(A \bu  + B \bu^{\otimes 2} \right)\otimes \bu
\nonumber\\
&=&\left( \mathds{1} \otimes A \right) \bu^{\otimes 2} + 
\left( \mathds{1} \otimes B \right) \bu^{\otimes 3}\nonumber\\&&+ \left( A \otimes \mathds{1} \right) \bu^{\otimes 2}
 +\left(  B \otimes \mathds{1} \right) \bu^{\otimes 3} \nonumber\\
 &=&\left( \mathds{1} \otimes A +A \otimes \mathds{1}\right) \bu^{\otimes 2} \nonumber\\&&+ 
\left( \mathds{1} \otimes B + B \otimes \mathds{1}\right) \bu^{\otimes 3} \,.
\eea
The general pattern thus becomes
\bea
\frac{\rmd \bu^{\otimes k}}{\rmd t} &=& \sum_{j=1}^k \underbrace{\bu \otimes \cdots \otimes \bu}_{1 \ldots j-1} 
\otimes \dot{\bu} \otimes \underbrace{\bu \otimes \cdots \otimes \bu}_{j+1 \cdots k} \nonumber\\&=&
\sum_{j=1}^k \underbrace{\bu \otimes \cdots \bu}_{1 \ldots j-1} 
\otimes \left(A \bu  + B \bu^{\otimes 2} \right)\otimes \underbrace{\bu \otimes \cdots \bu}_{j+1 \cdots k} 
\nonumber \\ &=&\left(\sum_{j=1}^kA^{(j,k)}\right)\bu^{\otimes k}+\left(\sum_{j=1}^kB^{(j,k)}\right)\bu^{\otimes(k+1)} \nonumber \\
&=&A_k \bu^{\otimes k}  + B_k \bu^{\otimes (k+1)}  \,,
\eea

Stacking the different tensor powers into one big vector, truncating the hierarchy, and padding each Carleman subspace to a common dimension gives us the matrix $\tilde{A}'+\tilde{B}'$.

\section{Derivation of the PMR expansion for the generator}\label{a:pmr}

Here we derive the PMR expansions for the padded matrices $\tilde{A}'$ and $\tilde{B}'$, thus establishing a PMR decomposition for the padded Carleman matrix.

Substituting the expansion of Eq.~\eqref{eq:Apmr} into Eq.~(\ref{eq:AkBk}) gives the PMR expansion of $A_k$:

\begin{equation}\label{eq:Akpmr}
\begin{aligned}
    A_k&=\frac{-2\nu}{a^2}\sum_{j=1}^k\mathds{1}^{\otimes k}+\sum_{j=1}^k\frac{\nu}{a^2}P_1^{(j,k)}+\sum_{j=1}^k\frac{\nu}{a^2}P_{-1}^{(j,k)}\\&=\frac{-2k\nu}{a^2}\mathds{1}^{\otimes k}+\sum_{j=1}^k\frac{\nu}{a^2}P_1^{(j,k)}+\sum_{j=1}^k\frac{\nu}{a^2}P_{-1}^{(j,k)},
\end{aligned}
\end{equation}
where $\mathds{1}^{\otimes k}$ is the $N^k\times N^k$ identity, and $P_{\pm1}^{(j,k)}$ denotes $P_{\pm1}$ tensored with $k-1$ identities, with $P_{\pm1}$ at the $j$-th position. Each $P_{\pm1}^{(j,k)}$ is itself a permutation matrix.

Using the padding of Eq.~\eqref{eq:Apadding} in Eq.~(\ref{eq:Akpmr}) then gives the PMR expansion for $A_k'$:

\begin{equation}
\begin{aligned}
    A_k'=\frac{-2k\nu}{a^2}\mathds{1}^{\otimes L}&+\frac{\nu}{a^2}\sum_{j=1}^k\left[P_1^{(j,k)}\otimes \mathds{1}^{\otimes L-k}\right]\\&+\frac{\nu}{a^2}\sum_{j=1}^k\left[P_{-1}^{(j,k)}\otimes \mathds{1}^{\otimes L-k}\right].
\end{aligned}
\end{equation}
Note that we can simplify $P_{\pm1}^{(j,k)}\otimes\mathds1^{\otimes L-k}$ to $P_{\pm1}^{(j,L)}$. Let $\Lambda_k$ denote the $L\times L$ matrix $\text{diag}(0,...,1,...,0)$, with the single $1$ at position $(k,k)$. Since each padded $A_k'$ has the common size $N^L\times N^L$, the full padded block-diagonal matrix $\tilde{A}'$ takes the form

\begin{equation}
\begin{aligned}
    \tilde{A}'&=\sum_{k=1}^L\Lambda_k\otimes A_k'\\&=\frac{-2\nu}{a^2}\sum_{k=1}^Lk\left(\Lambda_k\otimes\mathds{1}^{\otimes L}\right)\\&\quad+\frac{\nu}{a^2}\sum_{x=\pm1}\sum_{k=1}^L\sum_{j=1}^k(\Lambda_k\otimes\mathds{1}^{\otimes L})(\mathds{1}_L\otimes P_x^{(j,L)}),
\end{aligned}
\end{equation}
where $\mathds{1}_L$ is the $L\times L$ identity. This PMR can be simplified further by combining the diagonal masks, since the permutations are equal for all $k\geq j$. Rewriting the double sum indices $\sum_{k=1}^L\sum_{j=1}^k$ to $\sum_{j=1}^L\sum_{k=j}^L$ and absorbing the sum over $k$ into the diagonal, we have

\begin{equation}
\begin{aligned}
    \tilde{A}'&=\frac{-2\nu}{a^2}\sum_{k=1}^Lk\left(\Lambda_k\otimes\mathds{1}^{\otimes L}\right)\\&\quad+\frac{\nu}{a^2}\sum_{x=\pm1}\sum_{j=1}^L(\Lambda^{(j)}\otimes\mathds{1}^{\otimes L})(\mathds{1}_L\otimes P_x^{(j,L)}),
\end{aligned}
\end{equation}
where $\Lambda^{(j)}=\sum_{k=j}^L\Lambda_k$.

We take a similar route for $\tilde B'$. Substituting Eq.~\eqref{eq:Bpmr} into Eq.~(\ref{eq:AkBk}) gives
\begin{equation}
    B_k=(-1/2a)\sum_{j=1}^kQ_1^{(j,k)}+(1/2a)\sum_{j=1}^kQ_{-1}^{(j,k)},
\end{equation}
where $Q_{\pm1}^{(j,k)}$ denotes $Q_{\pm1}$ tensored with $k-1$ identities, placed at the $j$-th position.

In order to generalize Eq.~\eqref{eq:Bpadpmr} to $Q_{\pm1}^{(j,k)}$, note that $Q_{\pm1}^{(j,k)}$ contracts the adjacent registers $j$ and $j+1$, after which the surviving registers $j+2,\dots,k+1$ occupy the output slots $j+1,\dots,k$. We therefore first bring the registers into this final arrangement and only then perform the modular arithmetic. Let $R^{(j+1,k+1)}$ denote the cyclic left rotation of registers $j+1,\dots,k+1$ (implementable with $k-j$ register SWAPs, and trivial for $j=k$). Acting on a computational basis state, $R^{(j+1,k+1)}$ places the survivors $i_{j+2},\dots,i_{k+1}$ in slots $j+1,\dots,k$ and the consumed register $i_{j+1}$ in slot $k+1$; the adder $S_{\pm1}^{(j,k+1)}$ then maps slot $k+1$ to $i_{j+1}\ominus i_j\ominus(\pm1)$, and the projector onto $\ket{0}$ at slot $k+1$ enforces $i_{j+1}=i_j\oplus(\pm1)$. 

Comparing matrix elements, we obtain the exact operator identity
\begin{equation}
\begin{aligned}
    Q_{\pm1}^{(j,k)}\otimes \ket0=\left[\mathcal{F}_{\pm1}^{(j,k)}\otimes\ket{0}\bra{0}\right]S_{\pm1}^{(j,k+1)}R^{(j+1,k+1)}.
\end{aligned}
\end{equation}
with the second padding $\mathds1^{\otimes L-k-1}$ splitting between the diagonal mask and the permutation trivially, giving us the permutation and diagonal mask in Eq.~\eqref{eq:Bpermutation} and Eq.~\eqref{eq:Bdiagonal} respectively.

Returning to the full padded matrix, the block at the $k$-th off-diagonal position can be expressed as
\begin{equation}
\begin{pmatrix}
\ddots & & & & \\
& 0 & 0 & 0 & 0 \\
& 0 & 0 & B_k' & 0 \\
& 0 & 0 & 0 & 0 \\
& 0 & 0 & 0 & 0 \\
& & & & & \ddots
\end{pmatrix}=\begin{pmatrix}
\ddots & & & & \\
& 0 & 0 & 0 & 0 \\
& 0 & 0 & 1 & 0 \\
& 0 & 0 & 0 & 0 \\
& 0 & 0 & 0 & 0 \\
& & & & & \ddots
\end{pmatrix}\otimes B_k'
\end{equation}
where the first matrix is of dimension $L\times L$ and equals $\Lambda_kP_1^{(L)}$, with $P_1^{(L)}$ the $L\times L$ shift operator. 

Putting everything together, the full PMR expansion of the padded matrix $\tilde{B}'$ becomes:
\begin{equation}
\begin{aligned}
    &\tilde{B}'=\sum_{k=1}^{L-1}\Lambda_kP_1^{(L)}\otimes B_k'\\&=\frac{-1}{2a}\sum_{k=1}^{L-1}\sum_{j=1}^k(\Lambda_k\otimes \mathcal{F}_1^{(j,k,L)})(P_1^{(L)}\otimes S_1^{(j,k,L)})\\&+\frac{1}{2a}\sum_{k=1}^{L-1}\sum_{j=1}^k(\Lambda_k\otimes \mathcal{F}_{-1}^{(j,k,L)})(P_1^{(L)}\otimes S_{-1}^{(j,k,L)}).
\end{aligned}
\end{equation}

\section{Norm of $\tilde{A}'+\tilde{B}'$}\label{a2}
To bound the norm of $\tilde{A}'+\tilde{B}'$, we use the fact that the norm of a block matrix is bounded by the norm of the smaller matrix whose entries are the norms of the individual blocks, i.e. 
\begin{equation}
    \norm{\tilde{A}'+\tilde{B}'}\leq\left\|\begin{pmatrix}
        \norm{A_1'} & \norm{B_1'} & & \\
        & \norm{A_2'} & & \\
        & & \ddots & \\
        & & & \norm{B_{L-1}'}\\
        & &  & \norm{A_L'}
    \end{pmatrix}\right\|.
\end{equation}
It can be shown that $\norm{A_k'}=\norm{A_k}\leq k\norm{A}\leq 4k\nu/a^2$, and similarly $\norm{B_k'}=\norm{B_k}\leq k\norm{B}=k/(a\sqrt{2})$. For simplicity, let us denote $\norm{A}:=\chi_1$ and $\norm{B}=\chi_2$. We thus have:
\begin{equation}\label{b2}
\begin{aligned}
    \norm{\tilde{A}'+\tilde{B}'}&\leq\norm{\begin{pmatrix}
        \chi_1 & \chi_2 & & \\
        & 2\chi_1 & & \\
        & & \ddots & \\
        & & & (L-1)\chi_2\\
        & & & L\chi_1
    \end{pmatrix}}\\&=\norm{\text{diag}(1,2,\dots,L)\begin{pmatrix}
        \chi_1 & \chi_2 & & \\
        & \chi_1 & \ddots & \\
        & & \ddots & \chi_2\\
        & & & \chi_1\\
    \end{pmatrix}}\\&\leq L\norm{\chi_1\mathds{1}+\chi_2\begin{pmatrix}
        0 & 1 & & \\
        & 0 & \ddots & \\
        & & \ddots & 1\\
        & & & 0
    \end{pmatrix}}\\&\leq L(\chi_1+\chi_2)\\&=L\left(\frac{4\nu}{a^2}+\frac{1}{a\sqrt2}\right),
\end{aligned}
\end{equation}
where the second step uses submultiplicativity of the operator norm together with $\norm{\mathrm{diag}(1,2,\dots,L)}=L$ (the equality $\norm{\mathrm{diag}(1,\dots,L)\,T}=L\norm{T}$ does not hold in general, but the inequality suffices for the bound).

\section{Ensuring $G\succeq0$ for periodic boundary conditions for the Burgers' equation}\label{a3}
Recall that the matrix whose positive semidefiniteness we must verify is $-(\tilde{A}'+\tilde{B}')$. Thus $G=-\tilde{A}'-(\tilde{B}'+(\tilde{B}')^{\dagger})/2$. To establish $G\succeq0$, we compare the eigenvalues and norms of the individual matrices $\tilde{A}'$ and $\tilde{B}'$.

We begin by defining an arbitrary complex block vector $x=(x_1,x_2,\dots,x_L)$ where each block is of size $N^L$. Since the $A_k'$ are Hermitian (indeed real symmetric), we can write:
\begin{equation}
    x^\dagger Gx=-\sum_{k=1}^Lx_k^\dagger A_k'x_k-\sum_{k=1}^{L-1}\Re\!\left(x_k^\dagger B_k'x_{k+1}\right),
\end{equation}
where we have used $\frac12\big(x_k^\dagger B_k'x_{k+1}+x_{k+1}^\dagger {B_k'}^\dagger x_k\big)=\Re\big(x_k^\dagger B_k'x_{k+1}\big)$. Now by definition:
\begin{equation}
\begin{aligned}
    &x_k^\dagger A_k'x_k\geq\lambda_{\text{min}}(A_k')\norm{x_k}^2\\ \Rightarrow -&x_k^\dagger A_k'x_k\geq\lambda_{\text{min}}(-A_k')\norm{x_k}^2,
\end{aligned}
\end{equation}
and
\begin{equation}
\begin{aligned}
    \Re\!\left(x_k^\dagger B_k'x_{k+1}\right)&\leq\left|x_k^\dagger B_k'x_{k+1}\right|\\&\leq\norm{B_k'}\norm{x_k}\norm{x_{k+1}}\\ \Rightarrow-\Re\!\left(x_k^\dagger B_k'x_{k+1}\right)&\geq-\norm{B_k'}\norm{x_k}\norm{x_{k+1}}.
\end{aligned}
\end{equation}
Using $-2ab\geq-(a^2+b^2)$, we have:
\begin{equation}
\begin{aligned}
    -\Re\!&\left(x_k^\dagger B_k'x_{k+1}\right)\\&\geq-\frac{1}{2}(\norm{B_k'}\norm{x_k}^2+\norm{B_k'}\norm{x_{k+1}}^2).
\end{aligned}
\end{equation}
Thus we obtain the relation:
\begin{equation}
\begin{aligned}
    x^\dagger &Gx\\&\geq\sum_{k=1}^L\left(\lambda_{\text{min}}(-A_k')-\frac{\norm{B_k'}+\norm{B_{k-1}'}}{2}\right)\norm{x_k}^2,
\end{aligned}
\end{equation}
where we assume $B_0'=B_L'=0$. If the summand is non-negative for each $k$, then $x^\dagger Gx\geq0\ \forall \ x$, implying $G\succeq0$. We thus have the sufficient condition: 
\begin{equation}
    \lambda_{\text{min}}(-A_k')\geq\frac{\norm{B_k'}+\norm{B_{k-1}'}}{2}\ \ \forall \ \ k.
\end{equation}
The eigenvalues of $-A_k'$ are all non-negative. Unfortunately, for periodic boundary conditions, the minimum eigenvalue is 0, which fails to guarantee $G\succeq0$. This is remedied by adding a multiple of the identity that lifts the spectrum, i.e., by modifying $-(\tilde{A}'+\tilde{B}')$ to $-(\tilde{A}'+\tilde{B}')+\sigma\cdot \mathds1^{(L)}$, where $\mathds1^{(L)}$ is the $LN^L\times LN^L$ identity. 

Let us calculate the required eigenvalue shift $\sigma$. This constant will also determine how the gate costs are altered, due to the increased norm. We know that $\norm{B_k'}\leq k/(a\sqrt2)$ which implies
\begin{equation}
    \frac{\norm{B_k'}+\norm{B_{k-1}'}}{2}\leq k/(a\sqrt2).
\end{equation}
Thus if we set the constant $\sigma=L/(a\sqrt2)$, we get:
\begin{equation}
\begin{aligned}
    \lambda_{\text{min}}\left(-A_k'+\frac{L}{a\sqrt2}\cdot\mathds{1}\right)&\geq \frac{L}{a\sqrt2}\\&\geq\frac{\norm{B_k'}+\norm{B_{k-1}'}}{2}
\end{aligned}
\end{equation}
$\forall \ k$, which guarantees $G\succeq0$.

\section{LCHS state preparation}\label{app:Cprep}

Here we provide a method based on quantum rejection sampling ~\cite{10.1145/2090236.2090261} for the LCHS state preparation. It utilizes the structure of the quadrature coefficients to refine the elementary $\mathcal{O}(J)$ scaling to $\mathcal{O}(\sqrt K\log J)$ and additional dependence on the bits used for $\ket{k_j}$, thus resulting in an overall logarithmic scaling that can be absorbed into the soft-$\tilde{\mathcal{O}}$ notation in the final cost.

The aim is to prepare the normalized version of the state
\begin{equation}
    \sum_{j=0}^{J-1}\sqrt{c_j}\ket j=\sum_{m=-K/h_1}^{K/h_1-1}\sum_{q=0}^{Q_{GQ}-1}\sqrt{c_{q,m}}\ket{m,q},
\end{equation}
where $c_{q,m}=\frac{h_1}{2}w_qg(k_{q,m})$. Note that the factor $h_1/2$ will be cancelled upon normalization. Let $R=2K/h_1=J/Q_{GQ}$. We replace the physical register $m$ running from $-K/h_1$ to $K/h_1-1$ with a register running from 0 to $R-1$, and the physical register can be recovered with arithmetic. Start by preparing the state
\begin{equation}
    \frac{1}{\sqrt{R}}\sum_{m=0}^{R-1}\ket{m}\otimes\sum_{q=0}^{Q_{GQ}-1}\sqrt{\frac{w_q}{2}}\ket{q},
\end{equation}
which costs $\mathcal{O}(\log R+Q_{GQ})$. Note that this state is normalized because $\sum_qw_q=2$.

Next, we use $\ket{m,q}$ to load the nodes $k_{q,m}$ onto an additional register. This involves arithmetic, and only depends on the bit precision used for $k_{q,m}$. Using this register, we can use arithmetic to calculate $|g(k_{q,m})|$ and $\text{arg}(g(k_{q,m}))$, again depending only on the bit precision. 

Let $g_{\max}=\max_{|k|\leq K}|g(k)|=\mathcal{O}(1)$. We use a flag qubit to load the modulus as
\begin{equation}
    \ket k\ket 0\rightarrow\ket k\left(\sqrt{\frac{|g(k)|}{{g_{\max}}}}\ket 0+\sqrt{1-\frac{|g(k)|}{{g_{\max}}}}\ket1\right).
\end{equation}
The correctly flagged state will then be
\begin{equation}
    \frac{1}{\sqrt{2R{g_{\max}}}}\sum_{m,q}\sqrt{w_q|g(k_{q,m})|}\ket{m,q}\ket0.
\end{equation}
The success probability for the $\ket0$ flag is thus
\begin{equation}
    p_{\text{succ}}=\frac{\sum_{m,q}w_q|g(k_{q,m})|}{2R{g_{\max}}}=\frac{\norm{c}_1}{Rh_1{g_{\max}}}=\frac{\norm{c}_1}{2K{g_{\max}}}.
\end{equation}
In order to ensure the correct flag, we boost the probability through amplitude amplification, which uses
\begin{equation}
    \mathcal{O}(1/\sqrt{p_{\text{succ}}})=\mathcal{O}\left(\sqrt{K/\norm{c}_1}\right)=\mathcal{O}\left(\sqrt K\right)
\end{equation}
repetitions of the modulus state preparation. Here we have used $\norm{c}_1=\Theta(1)$. The final step is to add back the phase $e^{i\ \text{arg}(g(k_{q,m}))/2}$ (which is conjugated in the case of $C_{\text{prep}}^\dagger$), again depending on the bit precision. Thus the state preparation cost is
\begin{equation}
    \mathcal{O}\left(\sqrt K\left(\log R+Q_{GQ}+\text{poly}(b)\right)\right)=\tilde{\mathcal{O}}\left(\sqrt K\right),
\end{equation}
which is polylogarithmic in the relevant LCHS parameters. We can thus absorb $C_{\text{prep}}$ in the soft-$\tilde{\mathcal{O}}$ notation.

\section{Improvement to the postselection factor using rescaling}\label{a4}
To reduce the exponential postselection factor $e^{Lt/a\sqrt2}$ introduced by the stabilizing eigenvalue shift, we follow the rescaling scheme in Ref.~\cite{Costa_2025}. We note that rescaling $\bu$ is equivalent to re-optimizing the choice of the characteristic velocity $U_\star$ in the nondimensionalization of Sec.~\ref{sec:carleman}; the analysis below can thus be read as a principled optimization over that choice. Rescaling the vector $\bu\rightarrow\bu/\gamma$ consequently rescales the Carleman lifted vector $\tilde\bu$ as:
\begin{equation}
    \norm{\tilde\bu}^2=\sum_{k=1}^L\norm{\bu}^{2k}\rightarrow\sum_{k=1}^L\norm{\bu}^{2k}/\gamma^{2k},
\end{equation}
where $0<\gamma\leq1$. Let us call the rescaled norm $\norm{\tilde\bu_\gamma}$.  Define $w_k=\norm{\bu}^{2k}/\sum_k\norm{\bu}^{2k}$ such that $\sum_kw_k=1$. We know that $\gamma^{-2}\leq\gamma^{-2k}\leq\gamma^{-2L}$, thus $\gamma^{-1}\leq\norm{\tilde\bu_\gamma}/\norm{\tilde\bu}\leq\gamma^{-L}$. We can hence write
\begin{equation}
    \gamma^{L-1}\leq\frac{\norm{{\tilde\bu_{0 \gamma}}}/\norm{\tilde\bu_\gamma}}{\norm{\tilde\bu_0}/\norm{\tilde\bu}}\leq\gamma^{1-L}.
\end{equation}
Rescaling $\bu\rightarrow\bu/\gamma$ will rescale $B\rightarrow\gamma B$, leaving $A$ as it is. Consequently, the shift is rescaled to $\gamma L/a\sqrt2$. If we calculate the ratio of the postselection norm after and before rescaling, we get
\begin{equation}
    r\leq\frac{e^{(\gamma-1)Lt/a\sqrt2}}{\gamma^{L-1}}
\end{equation}.
Minimizing this ratio with respect to $\gamma$ gives us:
\begin{equation}
\begin{aligned}
    &r_{\min}\leq\frac{e^{L-1-L\chi}}{\left((L-1)/L\chi\right)^{L-1}}\\&\text{for }\gamma_{\min}=\frac{(L-1)}{L\chi}\ ,
\end{aligned}
\end{equation}
where we denote $\chi=t/a\sqrt2$ for brevity. In order for this minimum value to be valid, we must have the condition $\frac{(L-1)}{L\chi}\leq1$. Since $(L-1)/L<1$, a simpler condition is $\chi\geq1$. 

Using the property $(\frac{x}{x-1})^{x-1}<e$ for all $x>1$, we can show that the minimum rescaled postselection factor is bounded by:
\begin{equation}
    e^{L}\chi^{L-1}\norm{\tilde\bu_0}/\norm{\tilde\bu}.
\end{equation}
For fixed $L$, the dependence on $\chi$ (i.e. on $t$ and $N$) is now polynomial, which is an exponential improvement over the factor in Eq.~(\ref{eq:burgerscost}). Further, for fixed $\chi$, we can show
\begin{equation}
    \ln(r_{\min})\approx L(1-\chi)+(L-1)\ln\chi\ ,
\end{equation}
which for $\chi>1$ is always decreasing with $L$. Thus, the improvement ratio $r_{\min}$ will exponentially improve as we increase $L$. We note that this statement concerns the improvement \emph{ratio}: the absolute rescaled postselection factor $e^L\chi^{L-1}$ still grows with $L$, so increasing $L$ does not reduce the overall cost.

It should be noted that rescaling $\bu$ modifies the sector weights of the Carleman lifted vector $\ket{\tilde\bu}$, and hence the state to be prepared at $t=0$. Preparing $\ket{\tilde\bu_\gamma}$ is, however, the same class of state-preparation oracle as preparing $\ket{\tilde\bu_0}$, and is covered by the oracle assumption of Sec.~\ref{sec:carleman}; its cost is likewise absorbed into the soft-$\tilde{\mathcal{O}}$ of Eq.~(\ref{eq:burgerscost}).
\end{document}